\newif\ifeprint \eprinttrue                                          
\newif\ifsqueeze \squeezefalse 
\newif\ifanon \anonfalse 
\ifeprint \squeezetrue \anonfalse                                    
\PassOptionsToPackage{obeyspaces,spaces,hyphens}{url}                
\documentclass[rmp,twocolumn,floatfix,                               
  showkeys,letterpaper,amsmath]{revtex4}                             
\newcommand{\doi}[1]{\href                                           
  {https://doi.org/#1}{doi:\discretionary{}{}{}#1}}                  
\else                                                                
\documentclass[]{interact}
\usepackage{natbib}
\bibpunct[, ]{(}{)}{,}{a}{}{,}

\renewcommand{\doi}[1]{\href
  {https://doi.org/#1}{doi:\discretionary{}{}{}#1}}
\fi                                                                  
\usepackage{array,hyperref,graphicx,amssymb,orcidlink}
\ifanon\hypersetup{pdftitle=
  {On auxiliary latitudes},
  pdfauthor={Anonymous author(s)}}\else
\hypersetup{pdftitle=
  {On auxiliary latitudes},
  pdfauthor={C. F. F. Karney}}\fi
\renewcommand{\arraystretch}{1.25}
\newcommand{\gd}{\mathop{\mathrm{gd}}\nolimits}
\renewcommand{\d}{\mathrm d}
\usepackage{breakurl}
\ifpdf\def\urlalt#1#2{\burlalt{#2}{#1}}\else\let\urlalt=\burlalt\fi
\newcommand{\dlmf}[2]{\urlalt{http://dlmf.nist.gov/#2}{#1}}
\newcommand{\divdiff}{\mathord{\ooalign{$\triangle$\cr\hfil$\mid$\hfil\cr}}}
\def\figuredir{figures}

\begin{document}

\title{On auxiliary latitudes}

\ifeprint                                                   
\author{Charles F. F. Karney\,\orcidlink                    
  {0000-0002-5006-5836}}                                    
\email[Email: ]{charles.karney@sri.com}                     
\thanks{\href                                               
  {mailto:karney@alum.mit.edu}{karney@alum.mit.edu}.}       
\affiliation{SRI International,                             
  201 Washington Rd, Princeton, NJ  08540-6449, USA}        
\fi                                                         

\date{\today}
\begin{abstract}

The auxiliary latitudes are essential tools in cartography.  This
paper summarizes methods for converting between them with an emphasis on
providing full double-precision accuracy.  This includes series
expansions in the third flattening, where the truncation error is
precisely measured and where estimates of the radii of convergence are
given.  Also new formulas are given for computing the rectifying and
authalic latitudes with minimal roundoff error.

\end{abstract}

\ifeprint                                                            
\maketitle                                                           
\fi                                                                  
\section{Introduction}

The auxiliary latitudes, Table \ref{tab-aux}, map an ellipsoid of
revolution onto a sphere in various ways.  This allows an ellipsoidal
problem to be converted into a simpler spherical one.  They are used
extensively to tackle problems in geodesy and map projection.
For example, any equal area projection for the sphere can be converted
to an equal area projection on the ellipsoid by converting the latitude
on the ellipsoid to the authalic latitude on the sphere.  Similarly, the
formulas for rhumb lines on a sphere can be easily converted to apply to
an ellipsoid using the conformal and rectifying
latitudes \citep{karney-rhumb}.

Auxiliary latitudes have been part of the arsenal of geodetic techniques
for at least two centuries.  Why is another paper on this topic needed?
There are four reasons:
\begin{table}[tbp]
\caption{\label{tab-aux}The six auxiliary
latitudes $\phi$ thru $\xi$.  The isometric latitude, $\psi$, is not
counted as one of the auxiliary latitudes since it does not behave as an
angle (for example, it diverges at the poles).}
\begin{center}
\renewcommand{\arraystretch}{0.9}
\setlength{\extrarowheight}{1ex}
\begin{tabular}{@{\extracolsep{0.5em}}>{$}r<{$}l}
\hline\hline\noalign{\smallskip}
\eta& description\\
\noalign{\smallskip}\hline\noalign{\smallskip}
\phi & the (common, geodetic, geographic) latitude \\
\beta & the parametric (reduced) latitude \\
\theta & the geocentric latitude \\
\mu & the rectifying latitude \\
\chi & the conformal latitude \\
\xi & the authalic latitude \\
\noalign{\smallskip}\hline\noalign{\smallskip}
\eta, \zeta, \omega  & three generic auxiliary latitudes \\
\eta' & the complement of $\eta$, $\frac12\pi - \eta$, its colatitude \\
\psi & the isometric latitude, $\gd^{-1}\chi = \sinh^{-1}\tan\chi$\\
\noalign{\smallskip}
\hline\hline
\end{tabular}
\end{center}
\end{table}
\begin{itemize}
\item
We unambiguously document, Sec.~\ref{other-param}, the
superiority of expansions in the third flattening, $n$, compared to
expansions in the eccentricity squared, $e^2$.  The latter expansion
parameter is unfortunately still widely used, even though the truncation
errors are much larger.
\item
The paper catalogs the series expansions for all the conversions
between auxiliary latitudes, Appendix~\ref{series}.  These are also given
by \citet{orihuela13}.  We go substantially beyond this effort by
introducing a compact notation for the series, including the symbolic
manipulation code needed to produce the expansions, extending the
expansions to 40th order (see the supplementary data), and estimating
the radii of convergence for the series, Sec.~\ref{sec-radii}.
\item
In cases where the flattening is so large that auxiliary latitudes must
be computed by the defining equations, we reformulate the equations so
that they can be computed with little roundoff error,
Sec.~\ref{direct}.  Here we consider both the absolute error in the
latitudes and also the relative error in the tangents of the latitudes.
This allows accuracy to be maintained in a full range of
applications.
\item
An indispensable resource for map projections is \citet{snyder87}.
However, since its publication, geodetic measurements have become
increasingly precise, double-precision arithmetic has become
ubiquitous, and mapping packages have become embedded in complex software
systems; therefore, it is crucial that conversions between auxiliary
latitudes are carried out as accurately as possible.  This
paper updates the material on auxiliary latitudes in Snyder's \S3
to meet this goal.  The supplementary data includes C++ code
to implement the conversions either using the series expansions or via
direct evaluation, Sec.~\ref{implement}.
\end{itemize}

\ifanon We \else I \fi
do not give derivations of the basic formulas for the auxiliary
latitudes; these may be found in many textbooks and other sources,
e.g., \citet{adams21,snyder87,osborne13}.

\section{The auxiliary latitudes}\label{aux-sec}

The earth (or other celestial body) is usually approximated by an
ellipsoid of revolution with equatorial radius $a$ and polar semi-axis
$b$.  Several parameters may be used to express how eccentric the ellipse
is; these are listed in Table \ref{tab-params}.  All the formulas in
this paper apply equally to prolate and oblate ellipsoids; for prolate
ellipsoids, the eccentricity $e$ is pure imaginary; nonetheless,
expressions such as $e\tanh^{-1}(e\sin\phi)$ in Eq.~(\ref{psi-eq}), are
real.
\begin{table}[tbp]
\caption{\label{tab-params}
        Small parameters defining the shape of the ellipsoid.}
\begin{center}
\renewcommand{\arraystretch}{2.2}
\begin{tabular}{@{\extracolsep{1em}}r<{:}
  >{$\displaystyle}l<{$}}
\hline\hline\noalign{\smallskip}
the flattening          & f    =\frac{a-b}a = \frac{2n}{1+n} \\
2nd flattening          & f'   =\frac{a-b}b = \frac{2n}{1-n} \\
3rd flattening          & n    =\frac{a-b}{a+b} \\
the eccentricity squared& e^2  =\frac{a^2-b^2}{a^2} = \frac{4n}{(1+n)^2} \\
2nd eccentricity squared& e'^2 =\frac{a^2-b^2}{b^2} = \frac{4n}{(1-n)^2} \\
3rd eccentricity squared& e''^2=\frac{a^2-b^2}{a^2+b^2} = \frac{2n}{1+n^2} \\
\noalign{\smallskip}
\hline\hline\end{tabular}
\end{center}
\end{table}

The \emph{geographic} latitude $\phi$ (usually just referred to as the
``latitude'' but sometimes called the common or geodetic latitude) is
defined as the angle between the normal to the ellipsoid and the
equatorial plane.  It is the angle of the pole star above the horizon
and it is the only latitude that can be measured by traditional
navigational instruments.

If the ellipsoid is stretched into a sphere along the axis of rotation,
the resulting spherical latitude is the \emph{parametric} latitude,
\begin{equation}\label{beta-eq}
\beta = \tan^{-1}\bigl((1-f)\tan\phi\bigr),
\end{equation}
so called because the cartesian coordinates for points on the prime
meridian ellipse are
\begin{equation}
X = a\cos\beta; \quad Z = b\sin\beta.
\end{equation}
Some formulas in geodesy are simpler when expressed in terms of the
parametric latitude; for example, it plays a key role in solving for
geodesics by means of the auxiliary sphere \citep{bessel25-en}.  The
parametric latitude is also called the reduced latitude.

The \emph{geocentric} latitude,
\begin{equation}\label{theta-eq}
\theta = \tan^{-1}\bigl((1-f)^2\tan\phi\bigr),
\end{equation}
is self-explanatory.  This is used for example, when expressing the
magnetic field or the gravitational potential in terms of spherical
harmonics.

The other auxiliary latitudes are conveniently defined in terms of the
meridional and normal radii of curvature, $\rho$ and $\nu$,
\begin{align}
  \rho &= \frac{a^2 b^2}{(a^2 \cos^2\phi + b^2 \sin^2\phi)^{3/2}},
  \displaybreak[0]\\
  \nu &= \frac{a^2}{\sqrt{a^2 \cos^2\phi + b^2 \sin^2\phi}}.
\end{align}

The distance along the meridian ellipse measured from the equator to a
particular latitude is \citep[\S\dlmf{19.30(i)}{19.30.i}]{dlmf10}
\begin{align}
s &= \int_0 \rho \,\d\phi\notag\\
  &= \int_0 \sqrt{a^2 \sin^2\beta + b^2 \cos^2\beta} \,\d\beta\notag\\
  &= a\biggr( E(\phi, e)
   - \frac{e^2 \sin\phi \cos\phi}{\sqrt{1 - e^2 \sin^2\phi}}
  \biggr)\notag\\
  &= bE(\beta, ie'), \label{s-eq}
\end{align}
where $E(\phi,k)$ is the elliptic integral of the second kind
\citep[Eq.~\dlmf{19.2.E5}{19.30.5}]{dlmf10}.  The \emph{rectifying}
latitude is defined as
\begin{equation} \label{mu-eq}
  \mu = \frac\pi2 \frac s{s_p},
\end{equation}
where $s_p$ is the quarter meridian, the value of $s$ at the north pole.
Each degree of the rectifying latitude corresponds to the same distance
on the meridian.

The isometric latitude,
\begin{align}
\psi &= \int_0 \frac{\rho}{\nu\cos\phi} \,\d\phi\notag\\
     &= \gd^{-1} \phi - e \tanh^{-1}(e\sin\phi),\label{psi-eq}
\end{align}
is the vertical coordinate in the Mercator projection; here
\begin{equation}
\gd \psi = \tan^{-1}\sinh\psi \quad\text{and}\quad
\gd^{-1} \chi = \sinh^{-1}\tan\chi
\end{equation}
are the gudermannian and inverse gudermannian functions.
The Mercator projection is a conformal (angle preserving) projection of
the ellipsoid where circles of latitude and meridians are mapped to
straight lines.  If this projection is mapped conformally back onto a
sphere, the resulting latitude is the \emph{conformal} latitude,
\begin{equation}\label{chi-eq}
  \chi = \gd\psi;
\end{equation}
this allows the ellipsoid to be conformally mapped onto a sphere.

\citet[Eq.~7-7]{snyder87} writes
$\psi$ as
\begin{equation}
\psi =
\log\bigg[\tan\biggl(\frac\pi4 + \frac{\phi}2\biggr)
\biggl(\frac{1 - e \sin\phi}{1 + e \sin\phi}\biggr)^{e/2}\biggr].
\end{equation}
Although this is an equivalent expression, it is more cumbersome to
work with and results in larger roundoff errors compared to
Eq.~(\ref{psi-eq}); its use should be avoided.  Furthermore, the
alternative expressions for $\gd^{-1} \phi$,
\begin{equation}
\gd^{-1} \phi= \tanh^{-1}\sin\phi = 2\tanh^{-1}\tan{\textstyle\frac12}\phi,
\end{equation}
also lead to excessive roundoff errors.

The area between the equator and a circle of latitude is
\begin{align}
A &= 2\pi\int_0 \nu\cos\phi \,\rho \,\d\phi\notag\\
  &= \pi b^2\biggl(
    \frac{\tanh^{-1}(e\sin\phi)}e
    +\frac{\sin\phi}{1 - e^2\sin^2\phi} \biggr).
\label{Area-eq}
\end{align}
The \emph{authalic} latitude is defined in terms of $A$ by
\begin{equation}\label{xi-eq}
  \xi = \sin^{-1}\frac A{A_p},
\end{equation}
where $A_p$ (half the total area) is the value $A$ evaluated at the
north pole.  This provides an area-preserving mapping from the ellipsoid
to a sphere.

\begin{figure}[tbp]
\begin{center}
\includegraphics[scale=0.75]{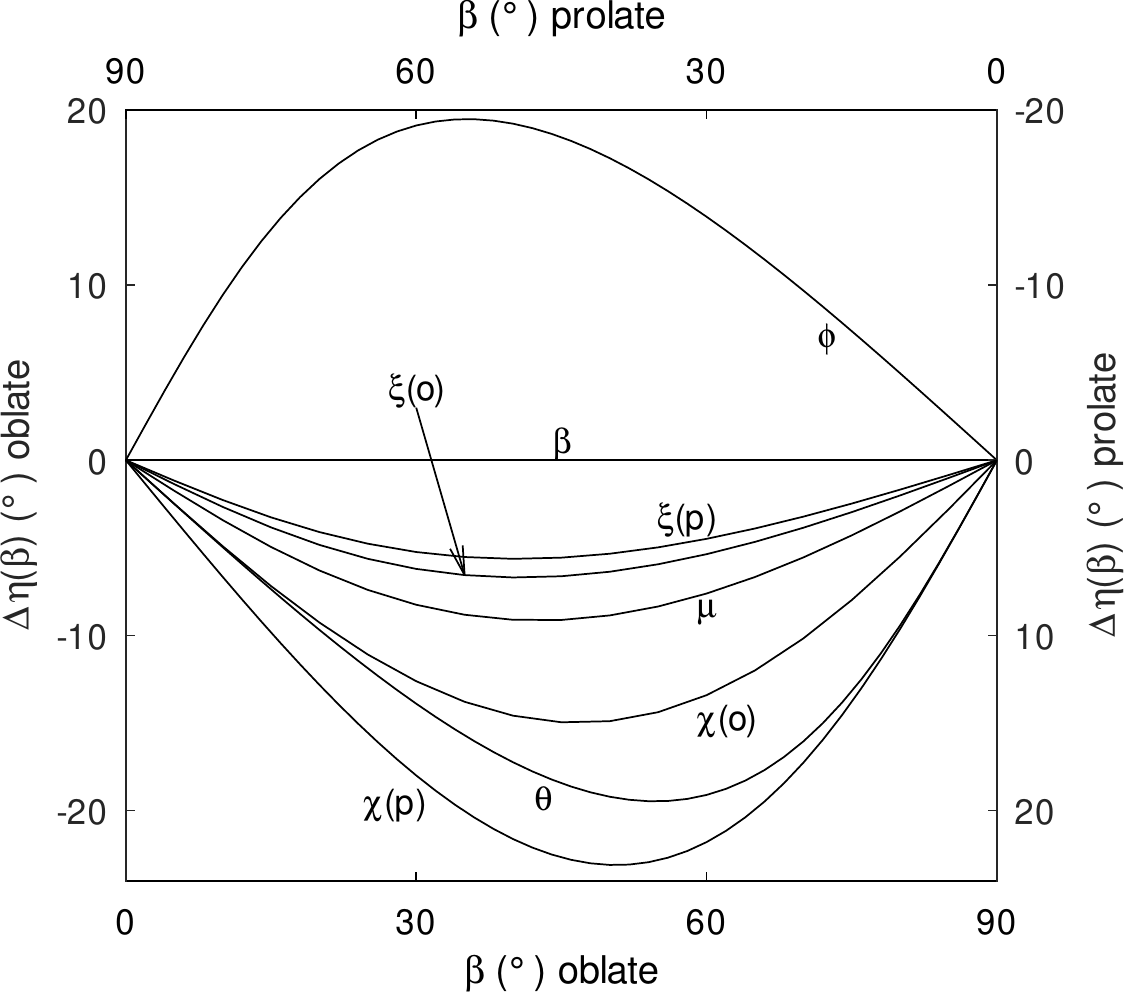}
\end{center}
\caption{\label{aux-fig}
Auxiliary latitudes for oblate and prolate ellipsoids with $n
= \pm\frac13$.  The quantities plotted are $\Delta\eta(\beta)
= \eta(\beta) - \beta$ as a function of $\beta$ for $\eta = \phi$,
$\beta$, $\theta$, $\mu$, $\chi$, and $\xi$.  For the oblate ellipsoid,
$n = \frac13$, $a/b = 2$, the scales are given by the bottom and left
axes; for the prolate ellipsoid, $n = -\frac13$, $a/b = \frac12$, the
scales are given by the \emph{reversed} top and right axes.  Because the
values of $\lvert n\rvert$ are the same, the curves for the oblate and
prolate cases coincide for the auxiliary latitudes, $\phi$, $\beta$,
$\theta$, and $\mu$.  For the other auxiliary latitudes, $\chi$ and
$\xi$, the curves for the oblate and prolate cases are labeled with (o)
and (p) respectively.}
\end{figure}%
The behavior of $\Delta\eta(\beta) = \eta(\beta) - \beta$ as a
function of $\beta$ is illustrated in Fig.~\ref{aux-fig} for an oblate
ellipsoid $a/b = 2$ and a prolate ellipsoid $a/b = \frac12$; here $\eta$
stands for any of the auxiliary latitudes.  This shows the symmetry in
the behavior for $\phi$, $\beta$, $\theta$, and $\mu$, when $a$ and $b$
are interchanged; this follows from the observation that the relations
between these auxiliary latitudes depend only on the properties of the
meridian ellipse (as opposed to the ellipsoid).  Also, because of
the symmetric relationship of $\phi$ and $\theta$ with $\beta$, the
curves, $\Delta\phi(\beta)$ and $\Delta\theta(\beta)$, are congruent.

\section{Series expansions}\label{sec-series}

The conversions between $\phi$, $\beta$, and $\theta$ are
straightforward.  However, the definition of $\mu$ involves a special
function and none of the expressions for $\mu$, $\chi$, or $\xi$ can be
inverted in closed form.  As a consequence, many authors have obtained
approximate expressions for the conversions based on the fact that the
flattening of the earth ellipsoid is small.  In this section, we examine
this approach.  Here we pick the third flattening $n$ as the expansion
parameter.  Some authors use $e^2$ instead; however, as we show in
Sec.~\ref{other-param}, this results in a much larger truncation error
for all the conversions.

Let's start by establishing a uniform and compact notation for the
series.  Take, for example, the series to convert $\phi$ to $\chi$ given
by \citet[Eq.~5.8]{krueger12},
\begin{align}
\chi(\phi) - \phi &\textstyle=
-\bigl(2n-\frac{2}{3}n^2-\frac{4}{3}n^3+\frac{82}{45}n^4\bigr)\sin2\phi
\notag\\&\textstyle\quad{}
+\bigl(\frac{5}{3}n^2-\frac{16}{15}n^3-\frac{13}{9}n^4\bigr)\sin4\phi
\notag\\&\textstyle\quad{}
-\bigl(\frac{26}{15}n^3-\frac{34}{21}n^4\bigr)\sin6\phi
\notag\\&\textstyle\quad{}
+\frac{1237}{630}n^4\sin8\phi + \ldots.
\end{align}
This can be written as
\begin{equation}
\Delta\chi(\phi) = \mathbf S^{(4)}(\zeta) \cdot
  \mathsf C_{\chi\phi}^{(4\times 4)} \cdot \mathbf P^{(4)}(n) + O(n^5),
\end{equation}
where
\begin{equation}
\mathbf S^{(L)}(\zeta) =
[\sin2\zeta, \sin4\zeta, \sin6\zeta, \ldots, \sin2L\zeta]
\end{equation}
is a row vector of length $L$,
\begin{equation}
\mathbf P^{(M)}(n) = [n, n^2, n^3, \ldots, n^M]^{\mathsf T}
\end{equation}
is a column vector of length $M$, and
\begin{equation}
\mathsf C_{\chi\phi}^{(4\times 4)}=
\begin{bmatrix}
-2&\frac{2}{3}&\frac{4}{3}&-\frac{82}{45}\\
0&\frac{5}{3}&-\frac{16}{15}&-\frac{13}{9}\\
0&0&-\frac{26}{15}&\frac{34}{21}\\
0&0&0&\frac{1237}{630}
\end{bmatrix}
\end{equation}
is an upper-triangular $4\times 4$ matrix.  All of the conversions from
one auxiliary latitude $\zeta$ to another $\eta$ follow the same
pattern,
\begin{equation}\label{gen-series}
  \Delta\eta(\zeta) = \mathbf S^{(L)}(\zeta) \cdot
  \mathsf C_{\eta\zeta}^{(L\times M)} \cdot \mathbf P^{(M)}(n) + O(n^{L+1}),
\end{equation}
allowing us to represent each series by a single matrix
$\mathsf C_{\eta\zeta}$.  (Hereafter we will usually omit the $(L\times
M)$ superscript on $\mathsf C_{\eta\zeta}$.)  We stipulate $M\ge L$; if
$M<L$, the last $L-M$ rows of $\mathsf C_{\eta\zeta}$ are zero---so it
may be replaced by a $M\times M$ matrix.  We can also write
\begin{align}
  \Delta\eta(\zeta) &= \mathbf S^{(L)}(\zeta) \cdot
  \mathbf F_{\eta\zeta}^{(L\times M)} + O(n^{L+1})\notag\\
  &=\sum_{l=1}^L F_{\eta\zeta,l}^{(L\times M)}\sin2l\zeta + O(n^{L+1}),
\end{align}
where
\begin{equation}\label{fourier}
\mathbf F_{\eta\zeta}^{(L\times M)} =
C_{\eta\zeta}^{(L\times M)} \cdot \mathbf P^{(M)}(n)
\end{equation}
is the length $L$
column vector of coefficients for the Fourier expansion of the periodic
function $\Delta\eta(\zeta)$ truncated at order $n^M$.  If
$M\rightarrow \infty$, then $\mathbf F_{\eta\zeta}^{(L\times \infty)}$
gives the first $L$ exact Fourier coefficients provided that the series
in $n$ converges (this question is addressed in Sec.~\ref{sec-radii}).
Usually we take $M = L$ as this defines the smallest matrix of
coefficients which gives accuracy to order $n^L$.

Using a computer algebra system, it's a simple matter to compute
$\mathsf C_{\eta\zeta}$ for all $\eta$ and $\zeta$.  \ifanon We \else I \fi
used \citet{maxima} which has embodied all the needed machinery (Taylor
series, trigonometric simplification, integration) since at least the
early 1970s.  It also has the virtue of being freely available.

The Maxima code to produce these expansions is relatively
straightforward and is included in the supplementary data for this paper.
This uses the result of \citet[p.~70]{delambre99} to give
$\mathsf C_{\beta\phi}$ and the expansion given by \citet{bessel25-en}
for $s$ in terms of $\beta$,
\begin{align}
  s &= \frac{a+b}2 \biggl(
  B_0 \beta + \sum_{l=1}^\infty \frac{B_l}l \sin2l\beta \biggl),
\displaybreak[0]\\
  B_l &= \sum_{\substack{j\ge l\\j-l\text{ even}}}
  \frac{(j+l-3)!!\,(j-l-3)!!}{(j+l)!!\,(j-l)!!} n^j,
\end{align}
as the starting point for $C_{\mu\beta}$.  (Here $j!!$ is the double
factorial extended to negative odd numbers so that $(-1)!!=1$ and
$(-3)!!=-1$.)  Series reversion was used to find $\mathsf C_{\zeta\eta}$
given $\mathsf C_{\eta\zeta}$.  One series can be substituted in another
to find $\mathsf C_{\eta\zeta}$ given $\mathsf C_{\eta\omega}$ and
$\mathsf C_{\omega\zeta}$.  In this way, all 30 non-trivial
$\mathsf C_{\eta\zeta}$ can be found.  These are listed for $M=L=6$ in
Appendix~\ref{series}, Eqs.~(\ref{C-beta-phi})--(\ref{C-chi-xi}).

The series for all the conversions between the 4 latitudes $\phi$,
$\beta$, $\theta$, and $\mu$ have alternating zeros in each row.  This
follows because interchanging $a$ and $b$, a symmetry exhibited in
Fig.~\ref{aux-fig}, changes the sign of $n$.  This property of the
expansions in $n$ was recognized by \citet{bessel25-en,helmert80-en};
earlier, \citet{euler55b} used $e''^2$ which has the same property.

Some of these expansions truncated to lower order can be found in the
literature.  In particular, \citet{krueger12} gives
$\mathsf C_{\chi\phi}$, his Eq.~5.8; $\mathsf C_{\phi\chi}$, Eq.~5.9;
$\mathsf C_{\chi\mu}$, Eq.~7.26*; and $\mathsf C_{\mu\chi}$, Eq.~8.41.
These series suffice to generalize the spherical transverse Mercator
projection to the ellipsoid.  Similarly \citet{helmert80-en}, in his
treatment of geodesics, gives $\mathsf C_{\mu\beta}$, his Eq.~5.5.7; and
$\mathsf C_{\beta\mu}$, Eq.~5.6.8.

\section{Truncation errors}\label{sec-trunc}

Modern geodetic libraries should strive to achieve full double-precision
accuracy.  This can be obtained at minimal cost and ensures that the
libraries continue to be relevant as geodetic measurements become more
precise.  As a starting point, consider the truncation errors for the
series expansions with $M=L=5$; as we shall see this gives accurate
results for terrestrial ellipsoids, exemplified by the International
Ellipsoid 1924, with $f = 1/297$.  The error
in evaluating $\Delta\eta(\zeta)$ is
\begin{equation}
\delta_{\eta\zeta}(\zeta) = \mathbf S(\zeta) \cdot
\bigl(\mathsf C_{\eta\zeta}^{(L\times L)}
    - \mathsf C_{\eta\zeta}^{(16\times16)}\bigr)
\cdot \mathbf P(n),
\end{equation}
where we assume that the results with $M=L=16$ are fully converged.
After the subtraction, the resulting matrix is
$\mathsf C_{\eta\zeta}^{(16\times16)}$ with the first $L=5$ columns set
to zero.  In this way, we can compute the truncation error using
double-precision arithmetic without the competing effects of roundoff.

\begin{table}[tbp]
\caption{\label{trunc-5} Absolute (a) and relative (b)
truncation errors of the series expansions in $n$ for $f = 1/297$,
truncated at $M=L=5$.  Here and in the following tables, the errors are
measured in ulps defined as $2^{-53}\,\mathrm{radian}$ for the absolute
errors and 1 part in $2^{53}$ for the relative errors.  The results are
given consistently to 2 significant figures.}
\begin{center}
(a)\\
\begin{tabular}{@{\extracolsep{1em}}
      >{$}c<{$} >{$}c<{$}>{$}c<{$}>{$}c<{$}>{$}c<{$}>{$}c<{$}>{$}c<{$}}
    \hline\hline\noalign{\smallskip}
    \eta\backslash\zeta
         &    \phi &   \beta &  \theta &     \mu &    \chi &     \xi \\
\noalign{\smallskip}\hline\noalign{\smallskip}
    \phi &       - &   0.035 &       6 &     2.3 &      17 &     1.1 \\
   \beta &   0.035 &       - &   0.035 &    0.35 &     3.4 &    0.15 \\
  \theta &       6 &   0.035 &       - &    0.23 &     1.9 &    0.11 \\
     \mu &    0.18 &  0.0024 &    0.67 &       - &    0.86 &   0.018 \\
    \chi &     2.1 &    0.11 &    0.56 &    0.11 &       - &   0.079 \\
     \xi &   0.086 &  0.0039 &    0.75 &   0.058 &     1.5 &       - \\
\noalign{\smallskip}
\hline\hline
\end{tabular}\\[2ex]
(b)\\
\begin{tabular}{@{\extracolsep{1em}}
      >{$}c<{$} >{$}c<{$}>{$}c<{$}>{$}c<{$}>{$}c<{$}>{$}c<{$}>{$}c<{$}}
    \hline\hline\noalign{\smallskip}
    \eta\backslash\zeta
         &    \phi &   \beta &  \theta &     \mu &    \chi &     \xi \\
\noalign{\smallskip}\hline\noalign{\smallskip}
    \phi &       - &    0.41 &      13 &     4.9 &      40 &     2.5 \\
   \beta &    0.41 &       - &    0.41 &    0.73 &      12 &     0.3 \\
  \theta &      13 &    0.41 &       - &    0.49 &     3.8 &    0.33 \\
     \mu &    0.66 &  0.0051 &     1.4 &       - &     4.6 &   0.037 \\
    \chi &     5.9 &    0.41 &     1.2 &    0.43 &       - &    0.33 \\
     \xi &    0.58 &  0.0086 &     1.6 &    0.12 &     6.9 &       - \\
\noalign{\smallskip}
\hline\hline
  \end{tabular}
\end{center}
\end{table}
\begin{table}[tbp]
\caption{\label{trunc-6} Absolute (a) and relative (b)
truncation errors of the series expansions in $n$ for $f = 1/150$,
truncated at $M=L=6$.  To obtain the corresponding values for other
(small) values of $f$, multiply the figures in this table by
$(150f)^7$; e.g., the truncation errors for $f=1/300$ are a factor $128$
smaller.}
\begin{center}
(a)\\
\begin{tabular}{@{\extracolsep{1em}}
      >{$}c<{$} >{$}c<{$}>{$}c<{$}>{$}c<{$}>{$}c<{$}>{$}c<{$}>{$}c<{$}}
    \hline\hline\noalign{\smallskip}
    \eta\backslash\zeta
         &    \phi &   \beta &  \theta &     \mu &    \chi &     \xi \\
\noalign{\smallskip}\hline\noalign{\smallskip}
    \phi &       - &   0.006 &     2.9 &    0.98 &       9 &    0.34 \\
   \beta &   0.006 &       - &   0.006 &    0.13 &     1.7 &    0.04 \\
  \theta &     2.9 &   0.006 &       - &   0.099 &    0.87 &    0.04 \\
     \mu &   0.037 & 0.00069 &    0.24 &       - &    0.31 &  0.0043 \\
    \chi &    0.78 &   0.018 &    0.18 &   0.022 &       - &   0.023 \\
     \xi &   0.015 & 0.00042 &    0.28 &   0.015 &     0.6 &       - \\
\noalign{\smallskip}
\hline\hline
\end{tabular}\\[2ex]
(b)\\
\begin{tabular}{@{\extracolsep{1em}}
      >{$}c<{$} >{$}c<{$}>{$}c<{$}>{$}c<{$}>{$}c<{$}>{$}c<{$}>{$}c<{$}}
    \hline\hline\noalign{\smallskip}
    \eta\backslash\zeta
         &    \phi &   \beta &  \theta &     \mu &    \chi &     \xi \\
\noalign{\smallskip}\hline\noalign{\smallskip}
    \phi &       - &   0.085 &     5.8 &       2 &      20 &    0.74 \\
   \beta &   0.085 &       - &   0.085 &    0.27 &     4.1 &    0.09 \\
  \theta &     5.8 &   0.085 &       - &     0.2 &     1.9 &   0.079 \\
     \mu &    0.13 &  0.0014 &    0.49 &       - &     1.5 &  0.0085 \\
    \chi &     1.7 &   0.085 &    0.36 &   0.055 &       - &   0.066 \\
     \xi &    0.12 & 0.00099 &    0.56 &   0.033 &     2.3 &       - \\
\noalign{\smallskip}
\hline\hline
\end{tabular}
\end{center}
\end{table}
The \emph{absolute} truncation error is just the maximum value of
$\lvert\delta_{\eta\zeta}(\zeta)\rvert$ as $\zeta$ is varied.
Table \ref{trunc-5}(a) gives these absolute errors for this case and
Table \ref{trunc-6}(a) gives the corresponding values for $M=L=6$ and
$f=1/150$; this is the value of $f$ for the SRMmax ellipsoid used by the
US National Geospatial-Intelligence Agency to ``stress test'' projection
libraries.  The errors are reported in ``ulps'', units in the last
place; for the purpose of measuring the absolute error of latitudes, we
define $1\,\mathrm{ulp} = 2^{-53}\,\mathrm{radian}$.  (Standard
double-precision hardware provides 53 bits of precision.)  Thus, values
less than $1\,\mathrm{ulp}$ mean that the truncation error is less than
the roundoff error and values in the range
$1\,\mathrm{ulp}$--$8\,\mathrm{ulp}$ mean that the truncation error is
comparable to the roundoff error.  We see that the series truncated at
$n^5$ are usually satisfactory for $f=1/297$; however, for production
use, it is wise to use the series truncated at $n^6$ (listed in
Appendix~\ref{series}) which give accurate results even for $f=1/150$.

In mapping applications, it is necessary to apply a more stringent error
metric, namely that the relative error in the latitude or the colatitude
be small.  For example, the radial coordinate of the polar stereographic
projection is proportional to $\rho_\chi = 2\tan\frac12\chi'$, where the
prime is used to signify the colatitude; in order to preserve accuracy
at the pole, the relative error in $\chi'$ must be small.  These
requirements can be met by representing the latitude by its tangent
because $\tan\eta$ can accurately represent angles extremely close to
$0$ and to $\pm\frac12\pi$.  Thus, $\rho_\chi$ can be expressed as
\begin{equation}
\rho_\chi
= 2\Bigl(\sqrt{1+\tan^2\chi}\pm \tan\chi\Bigr)^{\mp1},
\end{equation}
for $\chi\gtrless0$.
Similarly, the radial coordinate for the Lambert equal-area projection
is proportional to $\rho_\xi = 2\sin\frac12\xi'$ and this can be
evaluated as
\begin{equation}
\rho_\xi
= \sqrt{2\Bigl(\sqrt{1+\tan^2\xi}\pm \tan\xi\Bigr)^{\mp1}
\Big/\sqrt{1+\tan^2\xi}},
\end{equation}
for $\xi\gtrless0$.  This replaces the ill-conditioned expression,
$\sqrt{A_p-A}$, given by \citet[Eq.~24-23]{snyder87}.  To determine the
\emph{relative} error in $\tan\eta$, we multiply the absolute error by
$\d\log\tan\eta/\d\eta = 1/(\sin\eta\cos\eta)$.  This measure of the
relative error is given in Tables \ref{trunc-5}(b) and \ref{trunc-6}(b);
now, for the purposes of measuring the relative error, we define
$1\,\mathrm{ulp}$ to be 1 part in $2^{53}$.  Because the maximum value
of $\sin\eta\cos\eta$ is $\frac12$, the relative errors are all at least
twice as large as the absolute errors.

\begin{table}[tbp]
\caption{\label{trunc-fourier} Absolute (a) and relative (b)
truncation errors of the exact Fourier series ($M\rightarrow\infty$) for
$f = 1/150$, truncated at $L=6$.}
\begin{center}
(a)\\
\begin{tabular}{@{\extracolsep{\ifsqueeze0.7em\else1em\fi}}
      >{$}c<{$} >{$}c<{$}>{$}c<{$}>{$}c<{$}>{$}c<{$}>{$}c<{$}>{$}c<{$}}
    \hline\hline\noalign{\smallskip}
    \eta\backslash\zeta
         &    \phi &   \beta &  \theta &     \mu &    \chi &     \xi \\
\noalign{\smallskip}\hline\noalign{\smallskip}
    \phi &       - &   0.006 &    0.78 &    0.32 &     2.4 &    0.13 \\
   \beta &   0.006 &       - &   0.006 &   0.038 &    0.41 &   0.012 \\
  \theta &    0.78 &   0.006 &       - &   0.016 &    0.18 &  0.0046 \\
     \mu &   0.019 & 0.000097&   0.057 &       - &   0.046 & 0.00069 \\
    \chi &    0.15 & 0.000093&   0.031 &  0.0017 &       - & 0.00084 \\
     \xi &   0.011 & 0.00014 &   0.065 &  0.0031 &   0.095 &       - \\
\noalign{\smallskip}
\hline\hline
\end{tabular}\\[2ex]
(b)\\
\begin{tabular}{@{\extracolsep{1em}}
      >{$}c<{$} >{$}c<{$}>{$}c<{$}>{$}c<{$}>{$}c<{$}>{$}c<{$}>{$}c<{$}}
    \hline\hline\noalign{\smallskip}
    \eta\backslash\zeta
         &    \phi &   \beta &  \theta &     \mu &    \chi &     \xi \\
\noalign{\smallskip}\hline\noalign{\smallskip}
    \phi &       - &   0.085 &      11 &     4.4 &      33 &     1.8 \\
   \beta &   0.085 &       - &   0.085 &    0.53 &     5.7 &    0.17 \\
  \theta &      11 &   0.085 &       - &    0.23 &     2.5 &   0.065 \\
     \mu &    0.27 &  0.0014 &     0.8 &       - &    0.64 &  0.0096 \\
    \chi &     2.1 &  0.0013 &    0.43 &   0.024 &       - &   0.012 \\
     \xi &    0.15 &  0.0019 &     0.9 &   0.043 &     1.3 &       - \\
\noalign{\smallskip}
\hline\hline
\end{tabular}
\end{center}
\end{table}
\begin{figure}[tbp]
\begin{center}
\includegraphics[scale=0.75]{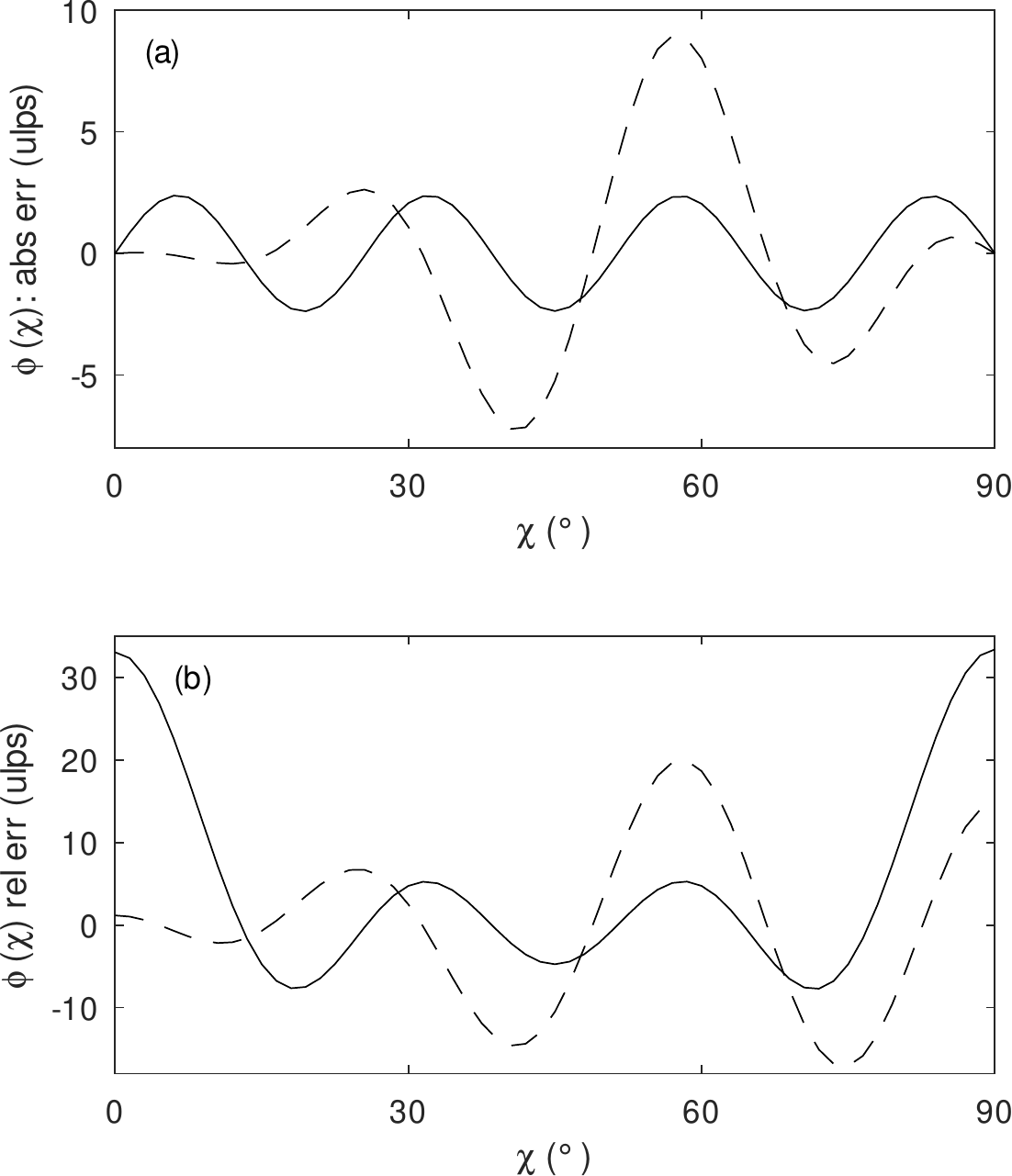}
\end{center}
\caption{\label{truncerr-fig}
Absolute (a) and relative (b) errors in computing $\phi(\chi)$ with the
expansions in $n$.  Here we have $f = 1/150$ and $L=6$ harmonics are
included; the solid (resp.~dashed) lines correspond to keeping
$M\rightarrow\infty$ (resp.~$M=6$) terms in the Taylor series expansions
for the Fourier coefficients.}
\end{figure}
So far, we have investigated the truncation error with $M=L$ which
corresponds to a consistent truncation to order $n^L$.  Since the vector
of Fourier coefficients $\mathbf F_{\eta\zeta}^{(L\times M)}$,
Eq.~(\ref{fourier}), can be
computed once the flattening of the ellipsoid is known, it might be
worthwhile to choose $M>L$ so that the vector more closely
approximates the exact Fourier coefficients.  For example, with $L=6$ and
$M=16$, we have a fully converged result for each Fourier coefficient for
$f = 1/150$.  The truncation error is now computed using
$\bigl(\mathsf C_{\eta\zeta}^{(L\times16)} -
\mathsf C_{\eta\zeta}^{(16\times16)}\bigr)$
which is the same as setting the first $L=6$ \emph{rows} of
$\mathsf C_{\eta\zeta}^{(16\times16)}$ to zero.
Table \ref{trunc-fourier} shows these truncation errors.  As
expected, all the absolute errors, Table \ref{trunc-fourier}(a), are
less than or equal to the corresponding values in
Table \ref{trunc-6}(a).  On the other hand, the relative errors in
Table \ref{trunc-fourier}(b) are, in many cases, greater than the
values in Table \ref{trunc-6}(b); adding more terms in the expansions
for the Fourier coefficients leads to a less accurate result (as judged
by the relative error).  This somewhat surprising result is explained by
Fig.~\ref{truncerr-fig} for the case of $\phi(\chi)$: The error in the
Fourier series with the exact coefficients, approximately given by the
next term in the Fourier series, oscillates sinusoidally as a function
of $\zeta$; in this way the Fourier series effectively minimizes the
maximum \emph{absolute} error; see the solid line in
Fig.~\ref{truncerr-fig}(a).  By the same token, this leads to an
increase in the \emph{relative} error near $\zeta = 0$ and $\frac12\pi$
where $\sin\eta\cos\eta$ is small; see the solid line in
Fig.~\ref{truncerr-fig}(b).  The dashed lines in this figure give the
errors with the Fourier coefficients truncated at $M=6$.

\section{Radii of convergence}\label{sec-radii}

The series expansions given here offer accurate conversions between
auxiliary latitudes for most geodetic applications.
As the eccentricity increases, it is possible to add additional terms to
the series so that accurate results are still obtained.  However, this
becomes increasingly fruitless for $\lvert f\rvert \gtrsim \frac1{10}$.
Generating the coefficients $\mathsf C_{\eta\zeta}^{(L\times L)}$
becomes impractical for $L > 20$, but more seriously, the series stop
converging for $\lvert n \rvert \ge r_{\eta\zeta}$, the ``radius of
convergence''.  For some conversions, e.g., $\mathsf C_{\beta\phi}$, the
radius of convergence is, from the behavior of the coefficients in
Eq.~(\ref{C-beta-phi}), $r_{\beta\phi} = 1$; i.e., the series
converges for all possible eccentricities; but for other conversions the
radius of convergence is considerably narrower.

\begin{figure}[tbp]
\begin{center}
\includegraphics[scale=0.75]{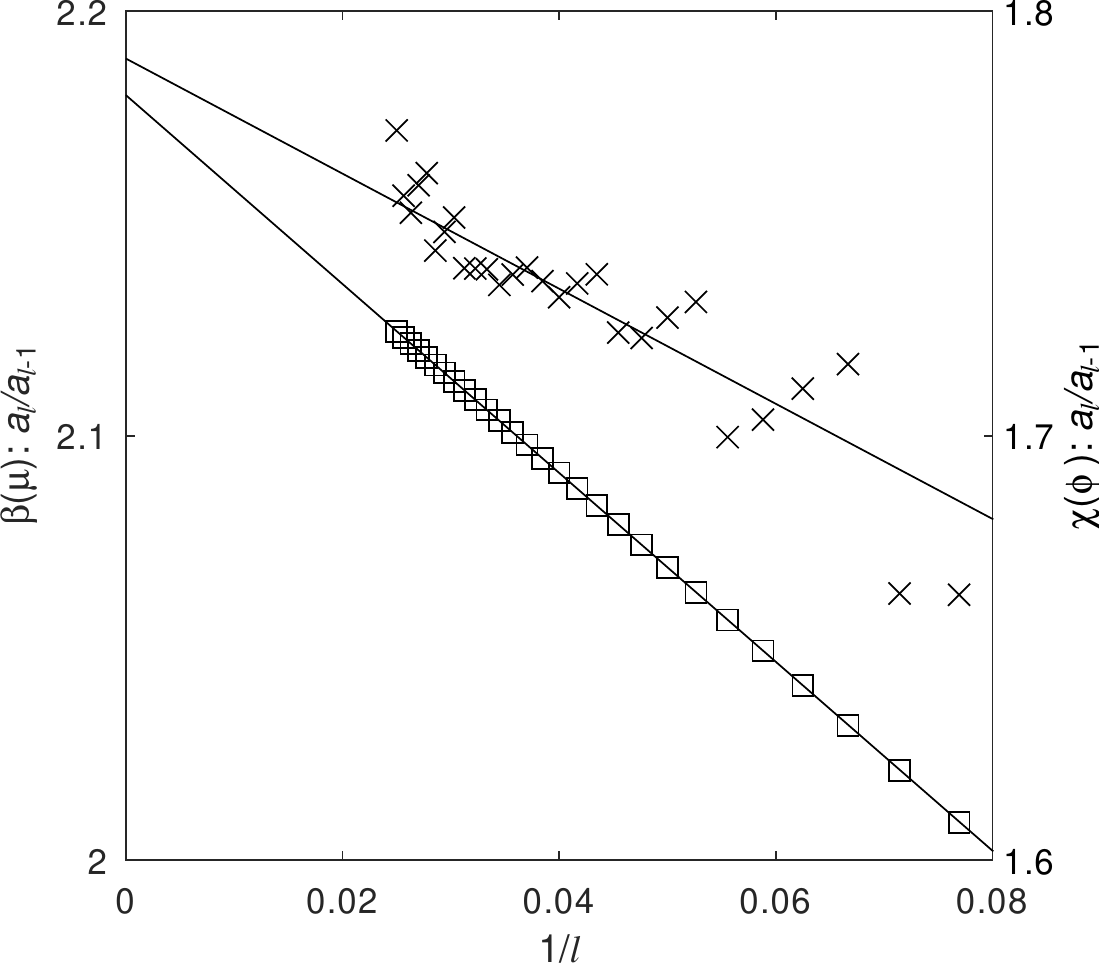}
\end{center}
\caption{\label{conv-fig}
Domb-Sykes plots to determine the convergence for the series expansions
in $n$ for $\beta(\mu)$ (squares, scale on left) and $\chi(\phi)$
(crosses, scale on right).  The lines give the least-squares linear fits
to the 20 points with $20<l\le40$.  The intercepts of the linear fits at
$1/l \rightarrow 0$ provide estimates of $1/r_{\beta\mu}$ and
$1/r_{\chi\phi}$.}
\end{figure}%
\begin{table*}[tbp]
\caption{\label{conv-tab}Radii of convergence, $r_{\eta\zeta}$, for the
expansions in $n$.  The values listed as ``1'' are exact.  The other
values are given with as many (up to 7) decimal places as are consistent
with the several fits to $a_l/a_{l-1}$.  Thus, the value $r_{\beta\mu} = 0.459$
means that the estimates all round to this result with 3 decimal places,
but not with 4 decimal places.  If the result isn't unique with 2
decimal places, then the range of values is given.  Thus, the certainty
of the estimate of $r_{\eta\zeta}$ is given by the number of decimal
places shown or by the range given.  In some cases, e.g.,
$r_{\mu\beta}$, the range includes impossible values greater than unity;
this just reflects the uncertainty of the fits.  The values for
$r_{\theta\phi}$ and $r_{\phi\theta}$ are very close to $\sqrt2 - 1$;
presumably this result could be established rigorously.}
\begin{center}
\begin{tabular}{@{\extracolsep{\ifsqueeze1em\else0.8em\fi}}
c cccccc} \hline\hline\noalign{\smallskip}
$\eta\backslash\zeta$ & $\phi$& $\beta$& $\theta$& $\mu$& $\chi$& $\xi$ \\
\noalign{\smallskip}\hline\noalign{\smallskip}
  $\phi$&       $-$& 1        & 0.4142136&    0.4587&0.32--0.35&      0.52 \\
 $\beta$& 1        &       $-$& 1        &     0.459&0.31--0.35&0.52--0.53 \\
$\theta$& 0.4142136& 1        &       $-$&    0.4590&0.33--0.34&0.51--0.53 \\
   $\mu$&      1.00&0.83--1.28&      0.41&       $-$&0.32--0.34&0.52--0.56 \\
  $\chi$&0.54--0.57&0.75--1.43&      0.41&      0.46&       $-$&0.51--0.57 \\
   $\xi$&1.00--1.01&0.99--1.00&      0.41&      0.46&0.33--0.35&       $-$ \\
\noalign{\smallskip}
\hline\hline
\end{tabular}
\end{center}
\end{table*}
A sufficient condition for convergence is that
$\lim_{L\rightarrow\infty}\sum_{l=1}^L a_l n^l$ is finite, where $a_l$
is the sum of the absolute values of the $l$th column of $\mathsf C$ (we
temporally drop the subscripts $\eta\zeta$).  The radius of convergence
is then $r = \lim_{l\rightarrow\infty} a_{l-1}/a_l$. We generated
$\mathsf C^{(40\times40)}$ using \citet{maxima}, and so could obtain
$a_l$ for $l \le 40$.  We can estimate $r$ using the \citet{domb57}
method, i.e., finding $r$ and $B$ which gives the least-squares linear
fit for $a_l/a_{l-1} \approx 1/r + B/l$.  We used the 20 values $l = 21$
through $40$ for the fit and we checked its soundness by using several
subsets of 10 out of the 20 points.  In some cases, the fit is
unambiguous, see the squares in Fig.~\ref{conv-fig} for the conversion
$\beta(\mu)$.  In others, the various fits yield different values of
$r$, see the crosses in the same figure for the conversion $\chi(\phi)$.
The results are summarized in Table \ref{conv-tab}.  The smallest radii
of convergence $r_{\eta\chi}$ are about $\frac13$; i.e., convergence for
$\frac12\lesssim b/a \lesssim 2$ or $-1\lesssim f \lesssim \frac12$.

\section{Evaluating the series}

Once the ellipsoid is selected, i.e., $n$ is known, then the vector of
Fourier coefficients $\mathbf F_{\eta\zeta}^{(L\times M)}$,
Eq.~(\ref{fourier}), can be evaluated.  Each coefficient is given as a
polynomial of order $M$ in $n$ and so can be evaluated rapidly and
accurately using the Horner method.  This evaluation should account for
the fact that the leading term in $F_{\eta\zeta,l}^{(L\times M)}$ is
$O(n^l)$.  Also, the polynomials for the conversions between $\phi$,
$\beta$, $\theta$, and $\mu$ are, because of the alternating zeros, in
$n^2$ rather than in $n$.

We are then left with the evaluation of the sums of the form
\begin{equation} \label{aux-clenshaw}
  \Delta\eta(\zeta) = \sum_{l=1}^L F_{\eta\zeta,l}^{(L\times M)}\sin2l\zeta.
\end{equation}
These are most conveniently evaluated using \citet{clenshaw55}
summation.  Although the method is well known, several software
libraries use more cumbersome methods for the summation.  So let us
review the method and reintroduce some less well known optimizations
that may be important in reducing roundoff errors.  We consider the more
general series
\begin{equation}
  p(\zeta) = \sum_{k=0}^{K-1} c_k f_k(\zeta),
\end{equation}
where
\begin{equation}
f_k(\zeta) = \sin\bigl(2(k+k_0)\zeta + \zeta_0\bigl).
\end{equation}
Equation \ref{aux-clenshaw} is the special case $k_0 = 1$, $\zeta_0 =
0$, $c_k = F_{\eta\zeta,k+1}^{(L\times M)}$ with $k = l - 1$.  The
general form of $f_k(\zeta)$ satisfies
\begin{equation}
f_{k-1}(\zeta) + f_{k+1}(\zeta) = 2x f_k(\zeta),
\end{equation}
where $x = (\cos\zeta + \sin\zeta)(\cos\zeta - \sin\zeta)$.
Clenshaw summation then proceeds as follows:
\begin{align}
  u_{K+1} &= u_{K} = 0, \displaybreak[0]\\
  u_k &= 2 x u_{k+1} - u_{k+2} + c_k, \quad\text{for $K > k \ge 0$},
\displaybreak[0]\\
  p(\zeta) &= u_0 f_0(\zeta) - u_1 f_{-1}(\zeta).
\end{align}

Note that the method involves no additional computation of trigonometric
functions beyond $\sin\zeta$ and $\cos\zeta$.  As with the Horner method,
the sum is computed starting with the highest-order (smallest) terms and
thus roundoff errors are minimized; in addition, the leading coefficient
in $p(\zeta)$, $c_0$, is $O(n)$ which further reduces the impact of
roundoff errors.

Evaluating the series for auxiliary latitudes, $f_k(\zeta)
= \sin\bigl(2(k+1)\zeta\bigl)$, we have $p(\zeta) =
2u_0\sin\zeta\cos\zeta$.  Because this product includes $\sin\zeta$ and
$\cos\zeta$, the \emph{relative} error due to roundoff is also small when
evaluating $\eta(\zeta)$ both near the equator ($\sin\zeta$ small)
and near a pole ($\cos\zeta$ small).

It is important to stress that, for the general case of Clenshaw
summation (with arbitrary $c_k$), the computation may be unstable if
$\sin\zeta$ or $\cos\zeta$ is small.  This is \emph{not} the case when
$c_k$ is strongly decreasing---the situation in our application.
However, for completeness, here is the modification
\citep[Reinsch, unpublished]{oliver77} needed in these cases:
\begin{align}
  x\mp1 &= \begin{cases}
  -2\sin^2\zeta,&\text{for upper sign, if $\lvert\sin\zeta\rvert$ small},\\
  2\cos^2\zeta,&\text{for lower sign, if $\lvert\cos\zeta\rvert$ small},
  \end{cases}
\displaybreak[0]\\
  d_K &= u_{K+1} = u_K = 0, \displaybreak[0]\\
  d_k &=2(x\mp1)u_{k+1} \pm d_{k+1} + c_k, \quad\text{for $K > k \ge 0$},
\displaybreak[0]\\
  u_k &= d_k \pm u_{k+1}, \quad\text{for $K > k \ge 0$},\displaybreak[0]\\
  p(\zeta) &= u_0 f_0(\zeta) - u_1 f_{-1}(\zeta).
\end{align}

\citet[\S3]{snyder87} writes the
trigonometric series, Eq.~(\ref{aux-clenshaw}), as a power series in $x$.
This also avoids multiple calls to trigonometric functions, but entails
manipulating the coefficients of the series.  Clenshaw summation offers
a more straightforward approach.

\section{Expanding in other small parameters}\label{other-param}

As mentioned at the beginning of Sec.~\ref{sec-series}, some authors use
the eccentricity squared, $e^2$, as the expansion parameter instead of
$n$.  It is rather simple to generate the resulting series.  Express $n$
in terms of $e^2$,
\begin{equation}
n = \frac{1 - \sqrt{1-e^2}}{1 + \sqrt{1-e^2}}= \frac14 e^2 + \frac18 e^4
+\frac5{64}e^6 + \ldots.
\end{equation}
We can then write
\begin{equation}
\mathbf P(n) = \mathsf T(n,e^2) \cdot \mathbf P(e^2),
\end{equation}
where $\mathsf T(n,e^2)$ is an $M\times M$ matrix whose $m$th row
consists of the coefficients of the expansion of $n^m$ in terms of
$e^2$.  Substituting in Eq.~(\ref{gen-series}), we obtain
\begin{equation}
  \Delta\eta(\zeta) = \mathbf S(\zeta) \cdot
  \mathsf C_{\eta\zeta} \cdot \mathsf T(n,e^2) \cdot
  \mathbf P(e^2) + O(n^{L+1}).
\end{equation}
Thus $\mathsf C_{\eta\zeta} \cdot \mathsf T(n,e^2)$ provides the matrix
of coefficients for the expansion in $e^2$.

\begin{table}[tbp]
\caption{\label{trunc-e2} Absolute (a) and relative (b)
truncation errors of the series expansions in $e^2$ for $f = 1/150$,
truncated at $M=L=6$.}
\begin{center}
(a)\\
\begin{tabular}{@{\extracolsep{1em}}
      >{$}c<{$} >{$}c<{$}>{$}c<{$}>{$}c<{$}>{$}c<{$}>{$}c<{$}>{$}c<{$}}
    \hline\hline\noalign{\smallskip}
    \eta\backslash\zeta
         &    \phi &   \beta &  \theta &     \mu &    \chi &     \xi \\
\noalign{\smallskip}\hline\noalign{\smallskip}
    \phi &       - &      28 &      82 &      59 &      71 &      53 \\
   \beta &      28 &       - &      28 &      12 &      16 &      10 \\
  \theta &      82 &      28 &       - &      23 &      23 &      23 \\
     \mu &      41 &     5.8 &      30 &       - &     1.2 &    0.54 \\
    \chi &      43 &     3.7 &      31 &    0.56 &       - &    0.53 \\
     \xi &      39 &     6.3 &      28 &    0.79 &     1.9 &       - \\
\noalign{\smallskip}
\hline\hline
\end{tabular}\\[2ex]
(b)\\
\begin{tabular}{@{\extracolsep{1em}}
      >{$}c<{$} >{$}c<{$}>{$}c<{$}>{$}c<{$}>{$}c<{$}>{$}c<{$}>{$}c<{$}}
    \hline\hline\noalign{\smallskip}
    \eta\backslash\zeta
         &    \phi &   \beta &  \theta &     \mu &    \chi &     \xi \\
\noalign{\smallskip}\hline\noalign{\smallskip}
    \phi &       - & 140     & 660     & 440     & 660     & 350     \\
   \beta & 140     &       - & 140     &      83 & 140     &      62 \\
  \theta & 660     & 140     &       - &      83 &      65 &      93 \\
     \mu & 230     &      19 &      68 &       - &     7.8 &     3.4 \\
    \chi &   300   &      11 &      68 &     2.3 &       - &     3.4 \\
     \xi &   200   &      22 &      67 &     5.8 &      20 &       - \\
\noalign{\smallskip}
\hline\hline
\end{tabular}
\end{center}
\end{table}
The transformation matrix $\mathsf T(n,e^2)$ is given by
Eq.~(\ref{T-e2}) and an example of the expansion in $e^2$ (for
converting from $\phi$ to $\mu$) is given by Eq.~(\ref{C-mu-phi-e2}).
Comparing this with Eq.~(\ref{C-mu-phi}), we note that the series in
terms of $e^2$ has lost the nice property of alternating zeros.  More
damning, the truncation errors for the expansions in $e^2$,
Table \ref{trunc-e2}, are \emph{all} much worse, in many cases by factors
in excess of 100, than for the corresponding expansions in $n$; compare
Table \ref{trunc-6}.

In light of this, the only reason to wish to convert a series in $n$
into the corresponding series in $e^2$ is to confirm that the truncation
error is larger and to cross-check the results with publications that
use this expansion.  However, the same technique can be profitably used
to convert an arbitrary series in $e^2$ to one in $n$: just multiply the
matrix of coefficients for the $e^2$ series by $\mathsf T(e^2,n)$
which is the matrix inverse of $\mathsf T(n,e^2)$, Eq.~(\ref{T-e2-inv}).

The procedure outlined here to obtain the series in $e^2$ can, of
course, be applied for any of the small parameters listed in
Table \ref{tab-params}.  The resulting matrices, $\mathsf T(n,\ldots)$,
are included in the supplementary data.  Besides $n$, the only parameter
with comparably small truncation errors is the third
eccentricity squared, $e''^2$, the choice of \citet{euler55b}.
Additionally, we remark that the series for converting between $\phi$
and $\theta$ is \emph{simpler} using $e''^2$ compared to $n$; for
example, we have $\mathsf C_{\theta\phi}\cdot\mathsf T(n,e''^2) =
\mathsf C_{\beta\phi}$, a diagonal matrix \citep[p.~13]{adams21}.

\section{Direct evaluation of latitudes}\label{direct}

The 6th-order series given in Appendix~\ref{series} provides full
double-precision accuracy for $\lvert f\rvert \le 1/150$.  For $f$
outside this range, we can add more terms to the series to maintain the
accuracy.  For example, the 8th-order series, $M=L=8$, are needed for
$\lvert f\rvert \le 1/50$.  However, even though extending the series to
40th order is feasible (see the supplementary data), going beyond 12th
order, say, is impractical: the number of coefficients become large
increasing the size of the program and the time to compute the
latitudes.  Furthermore, as we saw in Sec.~\ref{sec-radii}, some of the
series cease to converge for large $\lvert f\rvert$.

In those cases where using the series is not feasible, we can convert
between the auxiliary latitudes using the defining relations given in
Sec.~\ref{aux-sec}.  (Regardless of the value of $f$, this approach is
also recommended for the conversions between $\phi$, $\beta$, and
$\theta$.)  In order to ensure that the conversions maintain full
accuracy even at the equator and at the poles, we cast the relations in
terms of the tangents of the angles and demand that the \emph{relative}
roundoff errors are small.  The relations for $\beta$ and $\theta$ are
simple
\begin{align}
\tan\beta &= (1-f)\tan\phi,\label{tanbeta-eq}\displaybreak[0]\\
\tan\theta &= (1-f)^2\tan\phi.\label{tantheta-eq}
\end{align}

Given $\beta$, we can find $\mu$ using
Eqs.~(\ref{s-eq}) and (\ref{mu-eq}); however, these equations by themselves
will not yield good relative accuracy near the pole.  This is fixed by
computing not only $s$, the meridian distance from the equator, but
also $s' = s_p - s$, the distance from the pole, using
\begin{equation}
s' = a E(\beta', e).
\end{equation}
Then we have
\begin{equation}\label{tanmu-eq}
\tan\mu = \frac{\sin\bigl(\frac12\pi s/s_p\bigl)}
{\sin\bigl(\frac12\pi s'/s_p\bigl)}.
\end{equation}
(This assumes that $\beta\ge0$.  A similar equation can be written for
$\beta<0$.)  The elliptic functions can be evaluated with
\begin{align}
E(\zeta, k) &=
\sin\zeta
\bigl(
R_F(\cos^2\zeta, 1 - k^2\sin^2\zeta,1)\notag\\
&\quad{}-\textstyle\frac13 k^2\sin^2\zeta\,
R_D(\cos^2\zeta, 1 - k^2\sin^2\zeta, 1) \bigr),
\label{E1}
\end{align}
for $k^2 \le 0$, and
\begin{align}
E(\zeta, k) &=
\sin\zeta
\bigl(
k'^2\,R_F(\cos^2\zeta, k'^2 + k^2\cos^2\zeta,1)\notag\\
&\quad{}+\textstyle\frac13 k^2k'^2\sin^2\zeta\,
R_D(\cos^2\zeta, 1, k'^2 + k^2\cos^2\zeta)\notag\\
&\quad{}+k^2 \cos\zeta / \sqrt{k'^2 + k^2\cos^2\zeta} \bigr),
\label{E2}
\end{align}
where $k'^2 = 1 - k^2$, for $0 \le k^2 \le 1$.  These results assume
that $\lvert\zeta\rvert \le \frac12 \pi$.  The functions $R_F(x,y,z)$,
$R_D(x,y,z)$, and $R_G(x,y,z)$ (used below) are symmetric elliptic
integrals defined in
\citet[\S\dlmf{19.16(i)}{19.16.i}]{dlmf10} which can
be numerically evaluated to arbitrary precision using \citet{carlson95};
see also \citet[\S\dlmf{19.36(i)}{19.36.i}]{dlmf10}

The conditions on $k^2$ for Eqs.~(\ref{E1}) and (\ref{E2}) ensure that
the terms on the right hand sides are all positive thus minimizing the
roundoff error.  For an oblate (resp.~prolate) ellipsoid, we would
compute $s = bE(\beta, ie')$ with Eq.~(\ref{E1}) (resp.~Eq.~(\ref{E2}))
and $s' = bE(\beta', e)$ with Eq.~(\ref{E2}) (resp.~Eq.~(\ref{E1})).
The quarter meridian can be computed
with \citep[Eq.~\dlmf{19.25.1}{19.25.E1}]{dlmf10}
\begin{equation}
s_p = aE(e) = bE(ie') = 2 R_G(0,a^2,b^2);
\end{equation}
however, in practice, it is simpler to use $s_p = s + s'$.

Equations (\ref{E1}) and (\ref{E2}) are obtained from Eqs.~19.25.9 and
19.25.10
from \citet[\S\dlmf{19.25(i)}{19.25.i}]{dlmf10}.  However, the term
$\sin\zeta$ has been factored out of each equation.  This ensures good
relative accuracy when $\sin\zeta$ is small and this, in turn, ensures
high accuracy for Eq.~(\ref{tanmu-eq}) when either $\sin\beta$ or
$\cos\beta=\sin\beta'$ is small.

For treating the conformal latitude, we use the formulation
of \citet{karney11} combining Eqs.~(\ref{psi-eq}) and (\ref{chi-eq}) to
give
\begin{equation}\label{tanchi-eq}
\tan\chi = \tan\phi\sqrt{1+\sigma(\sin\phi)^2} -
  \sigma(\sin\phi)\sqrt{1+\tan^2\phi},
\end{equation}
where
\begin{equation}
\sigma(x) = \sinh\bigl(e\tanh^{-1}(e x)\bigr).
\end{equation}

Finally, we consider the equation for $\xi$, Eq.~(\ref{xi-eq}).  Since
this equation specifies just the sine of $\xi$, it results in a huge
\emph{absolute} error when evaluated numerically near a pole; this
results in a distance error of about $0.1\,\mathrm m$ on a terrestrial
ellipsoid.  The fix is to write the expression for $\cos\xi
= \sqrt{1-\sin^2\xi}$ and then manipulate the result to avoid large
cancellations.  This process gives
\begin{equation}\label{tanxi-eq}
\tan\xi = \frac{q(\sin\phi)}
{\cos\phi\sqrt{\divdiff[q](1,\sin\phi)\divdiff[q](1,-\sin\phi)}},
\end{equation}
where
\begin{equation}
q(x) = \frac{\tanh^{-1}(ex)}e + \frac x{1 - e^2x^2},
\end{equation}
and $\divdiff[q](x,y)$ is its ``divided difference'',
\begin{equation}
\divdiff[q](x,y) = \begin{cases}
\displaystyle \frac{q(y)-q(x)}{y-x}, & \text{for $xy < 0$},\\[2ex]
\displaystyle \frac{\d q(x)}{\d x}
= \frac2{(1-e^2x^2)^2}, & \text{for $y = x$},\\[2ex]
\displaystyle
\frac1{e(y-x)}\tanh^{-1}\frac{e(y-x)}{1-e^2xy}\\[2ex]
\quad\displaystyle{}
+\frac{1+e^2xy}{(1-e^2x^2)(1-e^2y^2)}, & \text{otherwise}.
\end{cases}
\end{equation}
The first of these expressions for $\divdiff[q](x,y)$ is the defining
relationship of the divided difference and this can be evaluated without
excessive roundoff if $x$ and $y$ have opposite signs; the second is the
result of taking the limit $y \rightarrow x$; and the final one is
obtained using the results of \citet{kahan99} and gives accurate results
even when $y \approx x$.

Equations (\ref{tanbeta-eq}), (\ref{tantheta-eq}), (\ref{tanmu-eq}),
(\ref{tanchi-eq}), and (\ref{tanxi-eq}), allow $\phi$ to be converted to
the other five auxiliary latitudes (with an intermediate conversion to
$\beta$ required for the conversion to $\mu$).  Equations
(\ref{tanbeta-eq}) and (\ref{tantheta-eq}) can be trivially inverted to
give $\phi$ in terms of $\beta$ or $\theta$.

The other equations cannot be inverted in closed form and for these we
use Newton's method to obtain an accurate result with just a few
iterations.  This method requires knowledge of the derivatives
$\d\tan\eta/\d\tan\phi$ which we list here:
\begin{align}
\frac{\d\tan\mu}{\d\tan\phi} &= \frac\pi2 \frac{(1-f)^2}{E(e)}
\frac{\cos^3\beta}{\cos\phi \cos^2\mu}, \displaybreak[0]\\
\frac{\d\tan\chi}{\d\tan\phi} &= (1-f)^2
\frac{\cos^2\beta}{\cos\chi\cos\phi},\displaybreak[0]\\
\frac{\d\tan\xi}{\d\tan\phi} &= \frac2{q(1)}
\frac{\cos^4\beta}{\cos^3\xi\cos\phi}.
\end{align}
We can easily take the limit, $\phi\rightarrow0$: merely replace the
cosine terms by unity.  In the limit, $\phi\rightarrow\frac12\pi$, we
have
\begin{align}
\frac{\d\tan\mu}{\d\tan\phi} &\rightarrow \frac2\pi (1-f){E(e)},
\displaybreak[0]\\
\frac{\d\tan\chi}{\d\tan\phi} &\rightarrow \sqrt{1+\sigma(1)^2}-\sigma(1),
\displaybreak[0]\\
\frac{\d\tan\xi}{\d\tan\phi} &\rightarrow (1-f)^2\sqrt{\frac{q(1)}2}.
\end{align}
Suitable starting guesses for $\tan\phi$ are $\tan\mu/({1-f})^{3/2}$,
$\tan\chi/({1-f})^2$, or $\tan\xi/({1-f})^{4/3}$; these are accurate guesses
for small $f$.

\section{Implementation}\label{implement}

The supplementary data for this paper includes a model C++
implementation of the methods described here.  The class AuxAngle is a
simple class that enables representation of an angle by a point in a
two-dimensional plane and this facilitates algorithms written in terms
of the tangent of the angle.  The class AuxLatitude implements the
conversions between all six auxiliary latitudes with the option of using
either the series expansions or the direct evaluation of the equations
for the latitudes.  The order of the series expansions can be selected
at compile time to be either $4$, $6$ (the default), or $8$.

The implementation allows the arithmetic to be carried out either at
standard double precision or using arbitrary precision arithmetic using
the MPREAL C++ interface \citep{mpreal} to the MPFR
library \citep{mpfr}.  This provides an alternative method of computing
the truncation errors given in Sec.~\ref{sec-trunc}: merely
evaluate, at high precision, the differences of the auxiliary
latitudes evaluated using the series method and the direct method.  This
serves as a confirmation that both methods are correctly coded.

However, the more interesting capability is the accurate measurement of
the roundoff errors for both the series and the direct evaluation methods.
This can be straightforwardly accomplished by taking the difference of
the results of the double-precision and high-precision calculations.
For the 6th-order series method, the absolute (resp.~relative) roundoff
error is about $2\,\mathrm{ulp}$ (resp.~$4\,\mathrm{ulp}$).  These
errors should be added to the truncation errors in Table~\ref{trunc-6},
scaled by $(150f)^7$.

Controlling the roundoff errors for the direct method is more
challenging.  We require, somewhat arbitrarily, an absolute error of no
more than $10\,\mathrm{ulp}$ (equivalent to about $7\,\mathrm{nm}$ in
the terrestrial ellipsoid) and a relative error of no more than
$30\,\mathrm{ulp}$; we limited the testing to $\lvert n\rvert \le 0.99$.
By this criterion, the formulas given in Sec.~\ref{direct} for
converting from $\phi$ to $\beta$, $\theta$, and $\mu$, are accurate for
the full range of $n$ tested.  However, conversions from $\phi$ to $\chi$
and $\xi$ are only accurate for $-0.69 \le n \le 0.40$.  We are able to
extend the region of accuracy for prolate ellipsoids by adapting the
formulas somewhat.  For example, we use
\begin{equation}
\frac{\tanh^{-1}(e\sin\phi)}e =
\begin{cases}
\displaystyle\frac{\tan^{-1}(\sqrt{-e^2}\sin\phi)}
{\sqrt{-e^2}}, & \text{for $f < 0$},\\
\sin\phi, & \text{for $f = 0$},\\[1ex]
\displaystyle\frac{\sinh^{-1}(e'\sin\beta)}{e}, & \text{for $f > 0$}.
\end{cases}
\end{equation}
The expressions for $f\le0$ are the obvious ones; however, for $f>0$,
putting the expression in terms of $\sinh^{-1}$ is much better
conditioned than the original expression with $\tanh^{-1}$, particularly
when $e\sin\phi$ is close to $1$.  This and other similar changes to the
formulas give accurate conversions from
$\phi$ to $\chi$ and $\xi$ for $-0.69 \le n \le 0.99$.  Refer to the
source files for details of these changes.

Newton's method for the conversions from $\mu$ and $\xi$ to $\phi$
converge in 7 iterations or less for $\lvert n\rvert \le 0.99$.  The
conversion from $\chi$ to $\phi$ is more problematic for prolate
ellipsoids, requiring more that 10 iterations for $n < -0.79$ and
failing to converge for $n < -0.81$.  The failure to converge is simple
to fix since the root finding problem has a single root.  With each
iteration, we update bounds on the value of $\tan\phi$.  If an
iteration places the updated value of $\tan\phi$ outside the bounds or
if the sign of the error oscillates, then the iteration is restarted
with $\tan\phi$ set to the geometric mean of the bounds.  This still
leads to an excessive number of iterations for extremely prolate
ellipsoids; this is cured by recasting the root finding problem in terms
of $\log\tan\phi$.  Again, details can be found in the source files for
this implementation.


\section{Conclusions}

This paper establishes accurate methods for converting between auxiliary
latitudes.  The time is long past when users of geodetic libraries
should be satisfied with ``millimeter precision''; nowadays libraries
should supply full double-precision accuracy and the methods provided
here achieve this at an acceptable computational cost.  By writing the
conversions in terms of the tangents of the latitudes so as to maintain
high relative accuracy, the same conversion routines can be used in a
wide range of applications.  For example, the conversions between $\phi$
and $\chi$ can be applied to the conformal projections in the different
aspects: cylindrical (Mercator), conic (Lambert conformal), and
azimuthal (polar stereographic).  Equal-area projections can likewise be
handled by a single pair of conversion routines between $\phi$ and
$\xi$.

The series given here offer some advantages over using the exact
conversion formulas: The roundoff errors are about two times smaller
and, provided $\lvert f \rvert \le \frac1{150}$, the truncation errors
with $M=L=6$ can be ignored.  Conversions between any auxiliary
latitudes are performed in one step.  By using Clenshaw summation, the
computation is usually faster compared to using the exact equations.  The
exception to this recommendation is for conversions between any of
$\phi$, $\beta$, and $\theta$, for which the exact formulas are so
simple that they should be used.

The series method generalizes simply to complex latitudes.  This is used
in the implementation of the transverse Mercator
projection \citep{krueger12} which entails converting between complex
conformal and rectifying latitudes using $C_{\mu\chi}$ and
$C_{\chi\mu}$, Eqs.~(\ref{C-chi-mu}) and (\ref{C-mu-chi}).

The paper establishes that series expansions using the third flattening
as the small parameter uniformly result in smaller truncation errors
compared to using the eccentricity.  Despite this, many projection
libraries adopt the formulas of \citet{adams21} and \citet{snyder87},
who use the inferior expansions for $\chi$ and $\xi$ in terms of the
eccentricity.  Section \ref{other-param} shows how a series in $e^2$ can
be converted to the equivalent series in $n$ by a simple matrix
multiplication (a purely arithmetic operation).

In cases where the series method is inapplicable, we reformulate the
defining equations for the latitudes so that they can be evaluated with
minimal roundoff error.  These provide accurate conversions for highly
prolate and oblate ellipsoids; for details, see Secs.~\ref{direct}
and \ref{implement}.

\ifanon
The methods for converting between $\phi$ and $\mu$ and $\xi$ using the
tangents of the latitudes, Eqs.~(\ref{tanmu-eq}) and (\ref{tanxi-eq}),
are new. \else
The method for converting between $\phi$ and $\chi$ using the tangents
of the latitudes, Eq.~(\ref{tanchi-eq}), and Newton's method is given
in \citet{karney11}.  The extensions of this method to
$\mu$ and $\xi$, Eqs.~(\ref{tanmu-eq}) and (\ref{tanxi-eq}), are new. \fi
The latter equation is particularly noteworthy since it cures the
catastrophic loss of accuracy in the usual formula for $\xi$ near the
poles.  This is also a good illustration of the power of divided
differences \citep{kahan99}; this technique can be used to improve the
numerical stability of the formulas in many geodetic applications.

\ifeprint \section*{Supplementary data} \fi                                                      

The following files are provided in \ifanon
\burl{https://www.dropbox.com/sh/hjdvrnu8rdgaynp/%
AAB_NJVrCpxaz90cI8V_cSg3a?dl=0}
\else
\begin{quote}
  \emph{Series expansions for converting
  \ifsqueeze \\ \fi
   between auxiliary latitudes},\\
  \doi{10.5281/zenodo.7382666},
\end{quote}
\fi
as supplementary data for this paper:
\begin{itemize}
\item {\tt auxlat.mac}:
the Maxima code used to obtain the series expansions given in the
appendix.  Instructions for using this code are included in the file.
\item {\tt auxvals40.mac}:
The matrices $\mathsf C$ and $\mathsf T$ for $M=L=40$ in a form suitable
for loading into maxima.
\item {\tt auxvals\{6,16,40\}.m}:
The matrices $\mathsf C$ and $\mathsf T$ in Octave/MATLAB notation for
$M=L=6$ and $16$ (using exact fractions) and $M=L=40$ (written as
floating-point numbers).  {\tt auxvals6.m} reproduces the results given
in Appendix~\ref{series} in ``machine-readable'' form. {\tt auxvals16.m}
was used to determine the truncation errors in Sec.~\ref{sec-trunc}.
{\tt auxvals40.m} was used to estimate the radii of convergence in
Sec.~\ref{sec-radii}.  The format of these files is sufficiently simple
that it can be easily adapted for any computer language.
\item {\tt AuxLatitude.hpp}, {\tt AuxLatitude.cpp}, and
{\tt example-AuxLatitude.cpp}: C++ implementations of the conversions
between auxiliary latitudes discussed in Sec.~\ref{implement} via the
classes AuxLatitude and AuxAngle.  \ifanon\else
An elaboration of these classes are included in Version 2.2 of
GeographicLib \citep{geographiclib22} where they are used to compute
rhumb lines \citep{karney-rhumb}. \fi
\end{itemize}


\bibliography{geod}
\appendix
\section{The series coefficients} \label{series}

\begin{table}[tbp]
  \caption{\label{eq-tab}
  Equation numbers for the coefficients $\mathsf C_{\eta\zeta}$.}
\begin{center}
\begin{tabular}{@{\extracolsep{1.5em}}c cccccc}
    \hline\hline\noalign{\smallskip}
$\eta\backslash\zeta$ & $\phi$ & $\beta$ & $\theta$ & $\mu$ & $\chi$ & $\xi$ \\
\noalign{\smallskip}\hline\noalign{\smallskip}
$\phi$   &$-$& \ref{C-beta-theta} & \ref{C-phi-theta} & \ref{C-phi-mu} & \ref{C-phi-chi} & \ref{C-phi-xi} \\
$\beta$  & \ref{C-beta-phi} &$-$& \ref{C-beta-theta} & \ref{C-beta-mu} & \ref{C-beta-chi} & \ref{C-beta-xi} \\
$\theta$ & \ref{C-theta-phi} & \ref{C-beta-phi} &$-$& \ref{C-theta-mu} & \ref{C-theta-chi} & \ref{C-theta-xi} \\
$\mu$    & \ref{C-mu-phi} & \ref{C-mu-beta} & \ref{C-mu-theta} &$-$& \ref{C-mu-chi} & \ref{C-mu-xi} \\
$\chi$   & \ref{C-chi-phi} & \ref{C-chi-beta} & \ref{C-chi-theta} & \ref{C-chi-mu} &$-$& \ref{C-chi-xi} \\
$\xi$    & \ref{C-xi-phi} & \ref{C-xi-beta} & \ref{C-xi-theta} & \ref{C-xi-mu} & \ref{C-xi-chi} &$-$\\
\noalign{\smallskip}
\hline\hline
\end{tabular}
\end{center}
\end{table}
Explicit formulas for $\mathsf C_{\eta\zeta}^{(6\times 6)}$ are given by
Eqs.~(\ref{C-beta-phi})--(\ref{C-chi-xi}) below.  These
duplicate the results of \citet{orihuela13} but with our more systematic
notation.  Table \ref{eq-tab} gives the equation numbers for all 30
matrices.  Since these are all upper-triangular matrices, the entries
below the main diagonal are left blank.  These equations are also given
in machine-readable form in the file {\tt auxvals6.m} included in the
supplementary data; this also includes the expansions for $M=L=40$.
{\ifsqueeze\small\fi
\begin{equation}\label{C-beta-phi}
\mathsf C_{\beta\phi}=
\mathsf C_{\theta\beta}=
\begin{bmatrix}
-1&0&0&0&0&0\\
&\frac{1}{2}&0&0&0&0\\
&&-\frac{1}{3}&0&0&0\\
&&&\frac{1}{4}&0&0\\
&&&&-\frac{1}{5}&0\\
&&&&&\frac{1}{6}
\end{bmatrix}
;\end{equation}
\begin{equation}\label{C-beta-theta}
\mathsf C_{\phi\beta}=
\mathsf C_{\beta\theta}=
\begin{bmatrix}
1&0&0&0&0&0\\
&\frac{1}{2}&0&0&0&0\\
&&\frac{1}{3}&0&0&0\\
&&&\frac{1}{4}&0&0\\
&&&&\frac{1}{5}&0\\
&&&&&\frac{1}{6}
\end{bmatrix}
;\end{equation}
\begin{equation}\label{C-theta-phi}
\mathsf C_{\theta\phi}=
\begin{bmatrix}
-2&0&2&0&-2&0\\
&2&0&-4&0&6\\
&&-\frac{8}{3}&0&8&0\\
&&&4&0&-16\\
&&&&-\frac{32}{5}&0\\
&&&&&\frac{32}{3}
\end{bmatrix}
;\end{equation}
\begin{equation}\label{C-phi-theta}
\mathsf C_{\phi\theta}=
\begin{bmatrix}
2&0&-2&0&2&0\\
&2&0&-4&0&6\\
&&\frac{8}{3}&0&-8&0\\
&&&4&0&-16\\
&&&&\frac{32}{5}&0\\
&&&&&\frac{32}{3}
\end{bmatrix}
;\end{equation}
\begin{equation}\label{C-mu-phi}
\mathsf C_{\mu\phi}=
\begin{bmatrix}
-\frac{3}{2}&0&\frac{9}{16}&0&-\frac{3}{32}&0\\
&\frac{15}{16}&0&-\frac{15}{32}&0&\frac{135}{2048}\\
&&-\frac{35}{48}&0&\frac{105}{256}&0\\
&&&\frac{315}{512}&0&-\frac{189}{512}\\
&&&&-\frac{693}{1280}&0\\
&&&&&\frac{1001}{2048}
\end{bmatrix}
;\end{equation}
\begin{equation}\label{C-phi-mu}
\mathsf C_{\phi\mu}=
\begin{bmatrix}
\frac{3}{2}&0&-\frac{27}{32}&0&\frac{269}{512}&0\\
&\frac{21}{16}&0&-\frac{55}{32}&0&\frac{6759}{4096}\\
&&\frac{151}{96}&0&-\frac{417}{128}&0\\
&&&\frac{1097}{512}&0&-\frac{15543}{2560}\\
&&&&\frac{8011}{2560}&0\\
&&&&&\frac{293393}{61440}
\end{bmatrix}
;\end{equation}
\begin{equation}\label{C-mu-beta}
\mathsf C_{\mu\beta}=
\begin{bmatrix}
-\frac{1}{2}&0&\frac{3}{16}&0&-\frac{1}{32}&0\\
&-\frac{1}{16}&0&\frac{1}{32}&0&-\frac{9}{2048}\\
&&-\frac{1}{48}&0&\frac{3}{256}&0\\
&&&-\frac{5}{512}&0&\frac{3}{512}\\
&&&&-\frac{7}{1280}&0\\
&&&&&-\frac{7}{2048}
\end{bmatrix}
;\end{equation}
\begin{equation}\label{C-beta-mu}
\mathsf C_{\beta\mu}=
\begin{bmatrix}
\frac{1}{2}&0&-\frac{9}{32}&0&\frac{205}{1536}&0\\
&\frac{5}{16}&0&-\frac{37}{96}&0&\frac{1335}{4096}\\
&&\frac{29}{96}&0&-\frac{75}{128}&0\\
&&&\frac{539}{1536}&0&-\frac{2391}{2560}\\
&&&&\frac{3467}{7680}&0\\
&&&&&\frac{38081}{61440}
\end{bmatrix}
;\end{equation}
\begin{equation}\label{C-mu-theta}
\mathsf C_{\mu\theta}=
\begin{bmatrix}
\frac{1}{2}&0&\frac{13}{16}&0&-\frac{15}{32}&0\\
&-\frac{1}{16}&0&\frac{33}{32}&0&-\frac{1673}{2048}\\
&&-\frac{5}{16}&0&\frac{349}{256}&0\\
&&&-\frac{261}{512}&0&\frac{963}{512}\\
&&&&-\frac{921}{1280}&0\\
&&&&&-\frac{6037}{6144}
\end{bmatrix}
;\end{equation}
\begin{equation}\label{C-theta-mu}
\mathsf C_{\theta\mu}=
\begin{bmatrix}
-\frac{1}{2}&0&-\frac{23}{32}&0&\frac{499}{1536}&0\\
&\frac{5}{16}&0&-\frac{5}{96}&0&\frac{6565}{12288}\\
&&\frac{1}{32}&0&-\frac{77}{128}&0\\
&&&\frac{283}{1536}&0&-\frac{4037}{7680}\\
&&&&\frac{1301}{7680}&0\\
&&&&&\frac{17089}{61440}
\end{bmatrix}
;\end{equation}
\begin{equation}\label{C-chi-phi}
\mathsf C_{\chi\phi}=
\begin{bmatrix}
-2&\frac{2}{3}&\frac{4}{3}&-\frac{82}{45}&\frac{32}{45}&\frac{4642}{4725}\\
&\frac{5}{3}&-\frac{16}{15}&-\frac{13}{9}&\frac{904}{315}&-\frac{1522}{945}\\
&&-\frac{26}{15}&\frac{34}{21}&\frac{8}{5}&-\frac{12686}{2835}\\
&&&\frac{1237}{630}&-\frac{12}{5}&-\frac{24832}{14175}\\
&&&&-\frac{734}{315}&\frac{109598}{31185}\\
&&&&&\frac{444337}{155925}
\end{bmatrix}
;\end{equation}
\begin{equation}\label{C-phi-chi}
\mathsf C_{\phi\chi}=
\begin{bmatrix}
2&-\frac{2}{3}&-2&\frac{116}{45}&\frac{26}{45}&-\frac{2854}{675}\\
&\frac{7}{3}&-\frac{8}{5}&-\frac{227}{45}&\frac{2704}{315}&\frac{2323}{945}\\
&&\frac{56}{15}&-\frac{136}{35}&-\frac{1262}{105}&\frac{73814}{2835}\\
&&&\frac{4279}{630}&-\frac{332}{35}&-\frac{399572}{14175}\\
&&&&\frac{4174}{315}&-\frac{144838}{6237}\\
&&&&&\frac{601676}{22275}
\end{bmatrix}
;\end{equation}
\begin{equation}\label{C-chi-beta}
\mathsf C_{\chi\beta}=
\begin{bmatrix}
-1&\frac{2}{3}&0&-\frac{16}{45}&\frac{2}{5}&-\frac{998}{4725}\\
&\frac{1}{6}&-\frac{2}{5}&\frac{19}{45}&-\frac{22}{105}&-\frac{2}{27}\\
&&-\frac{1}{15}&\frac{16}{105}&-\frac{22}{105}&\frac{116}{567}\\
&&&\frac{17}{1260}&-\frac{8}{105}&\frac{2123}{14175}\\
&&&&-\frac{1}{105}&\frac{128}{4455}\\
&&&&&\frac{149}{311850}
\end{bmatrix}
;\end{equation}
\begin{equation}\label{C-beta-chi}
\mathsf C_{\beta\chi}=
\begin{bmatrix}
1&-\frac{2}{3}&-\frac{1}{3}&\frac{38}{45}&-\frac{1}{3}&-\frac{3118}{4725}\\
&\frac{5}{6}&-\frac{14}{15}&-\frac{7}{9}&\frac{50}{21}&-\frac{247}{270}\\
&&\frac{16}{15}&-\frac{34}{21}&-\frac{5}{3}&\frac{17564}{2835}\\
&&&\frac{2069}{1260}&-\frac{28}{9}&-\frac{49877}{14175}\\
&&&&\frac{883}{315}&-\frac{28244}{4455}\\
&&&&&\frac{797222}{155925}
\end{bmatrix}
;\end{equation}
\begin{equation}\label{C-chi-theta}
\mathsf C_{\chi\theta}=
\begin{bmatrix}
0&\frac{2}{3}&\frac{2}{3}&-\frac{2}{9}&-\frac{14}{45}&\frac{1042}{4725}\\
&-\frac{1}{3}&\frac{4}{15}&\frac{43}{45}&-\frac{4}{45}&-\frac{712}{945}\\
&&-\frac{2}{5}&\frac{2}{105}&\frac{124}{105}&\frac{274}{2835}\\
&&&-\frac{55}{126}&-\frac{16}{105}&\frac{21068}{14175}\\
&&&&-\frac{22}{45}&-\frac{9202}{31185}\\
&&&&&-\frac{90263}{155925}
\end{bmatrix}
;\end{equation}
\begin{equation}\label{C-theta-chi}
\mathsf C_{\theta\chi}=
\begin{bmatrix}
0&-\frac{2}{3}&-\frac{2}{3}&\frac{4}{9}&\frac{2}{9}&-\frac{3658}{4725}\\
&\frac{1}{3}&-\frac{4}{15}&-\frac{23}{45}&\frac{68}{45}&\frac{61}{135}\\
&&\frac{2}{5}&-\frac{24}{35}&-\frac{46}{35}&\frac{9446}{2835}\\
&&&\frac{83}{126}&-\frac{80}{63}&-\frac{34712}{14175}\\
&&&&\frac{52}{45}&-\frac{2362}{891}\\
&&&&&\frac{335882}{155925}
\end{bmatrix}
;\end{equation}
\begin{equation}\label{C-chi-mu}
\mathsf C_{\chi\mu}=
\begin{bmatrix}
-\frac{1}{2}&\frac{2}{3}&-\frac{37}{96}&\frac{1}{360}&\frac{81}{512}&-\frac{96199}{604800}\\
&-\frac{1}{48}&-\frac{1}{15}&\frac{437}{1440}&-\frac{46}{105}&\frac{1118711}{3870720}\\
&&-\frac{17}{480}&\frac{37}{840}&\frac{209}{4480}&-\frac{5569}{90720}\\
&&&-\frac{4397}{161280}&\frac{11}{504}&\frac{830251}{7257600}\\
&&&&-\frac{4583}{161280}&\frac{108847}{3991680}\\
&&&&&-\frac{20648693}{638668800}
\end{bmatrix}
;\end{equation}
\begin{equation}\label{C-mu-chi}
\mathsf C_{\mu\chi}=
\begin{bmatrix}
\frac{1}{2}&-\frac{2}{3}&\frac{5}{16}&\frac{41}{180}&-\frac{127}{288}&\frac{7891}{37800}\\
&\frac{13}{48}&-\frac{3}{5}&\frac{557}{1440}&\frac{281}{630}&-\frac{1983433}{1935360}\\
&&\frac{61}{240}&-\frac{103}{140}&\frac{15061}{26880}&\frac{167603}{181440}\\
&&&\frac{49561}{161280}&-\frac{179}{168}&\frac{6601661}{7257600}\\
&&&&\frac{34729}{80640}&-\frac{3418889}{1995840}\\
&&&&&\frac{212378941}{319334400}
\end{bmatrix}
;\end{equation}
\begin{equation}\label{C-xi-phi}
\mathsf C_{\xi\phi}=
\begin{bmatrix}
-\frac{4}{3}&-\frac{4}{45}&\frac{88}{315}&\frac{538}{4725}&\frac{20824}{467775}&-\frac{44732}{2837835}\\
&\frac{34}{45}&\frac{8}{105}&-\frac{2482}{14175}&-\frac{37192}{467775}&-\frac{12467764}{212837625}\\
&&-\frac{1532}{2835}&-\frac{898}{14175}&\frac{54968}{467775}&\frac{100320856}{1915538625}\\
&&&\frac{6007}{14175}&\frac{24496}{467775}&-\frac{5884124}{70945875}\\
&&&&-\frac{23356}{66825}&-\frac{839792}{19348875}\\
&&&&&\frac{570284222}{1915538625}
\end{bmatrix}
;\end{equation}
\begin{equation}\label{C-phi-xi}
\mathsf C_{\phi\xi}=
\begin{bmatrix}
\frac{4}{3}&\frac{4}{45}&-\frac{16}{35}&-\frac{2582}{14175}&\frac{60136}{467775}&\frac{28112932}{212837625}\\
&\frac{46}{45}&\frac{152}{945}&-\frac{11966}{14175}&-\frac{21016}{51975}&\frac{251310128}{638512875}\\
&&\frac{3044}{2835}&\frac{3802}{14175}&-\frac{94388}{66825}&-\frac{8797648}{10945935}\\
&&&\frac{6059}{4725}&\frac{41072}{93555}&-\frac{1472637812}{638512875}\\
&&&&\frac{768272}{467775}&\frac{455935736}{638512875}\\
&&&&&\frac{4210684958}{1915538625}
\end{bmatrix}
;\end{equation}
\begin{equation}\label{C-xi-beta}
\mathsf C_{\xi\beta}=
\begin{bmatrix}
-\frac{1}{3}&-\frac{4}{45}&\frac{32}{315}&\frac{34}{675}&\frac{2476}{467775}&-\frac{70496}{8513505}\\
&-\frac{7}{90}&-\frac{4}{315}&\frac{74}{2025}&\frac{3992}{467775}&\frac{53836}{212837625}\\
&&-\frac{83}{2835}&\frac{2}{14175}&\frac{7052}{467775}&-\frac{661844}{1915538625}\\
&&&-\frac{797}{56700}&\frac{934}{467775}&\frac{1425778}{212837625}\\
&&&&-\frac{3673}{467775}&\frac{390088}{212837625}\\
&&&&&-\frac{18623681}{3831077250}
\end{bmatrix}
;\end{equation}
\begin{equation}\label{C-beta-xi}
\mathsf C_{\beta\xi}=
\begin{bmatrix}
\frac{1}{3}&\frac{4}{45}&-\frac{46}{315}&-\frac{1082}{14175}&\frac{11824}{467775}&\frac{7947332}{212837625}\\
&\frac{17}{90}&\frac{68}{945}&-\frac{338}{2025}&-\frac{16672}{155925}&\frac{39946703}{638512875}\\
&&\frac{461}{2835}&\frac{1102}{14175}&-\frac{101069}{467775}&-\frac{255454}{1563705}\\
&&&\frac{3161}{18900}&\frac{1786}{18711}&-\frac{189032762}{638512875}\\
&&&&\frac{88868}{467775}&\frac{80274086}{638512875}\\
&&&&&\frac{880980241}{3831077250}
\end{bmatrix}
;\end{equation}
\begin{equation}\label{C-xi-theta}
\mathsf C_{\xi\theta}=
\begin{bmatrix}
\frac{2}{3}&-\frac{4}{45}&\frac{62}{105}&\frac{778}{4725}&-\frac{193082}{467775}&-\frac{4286228}{42567525}\\
&\frac{4}{45}&-\frac{32}{315}&\frac{12338}{14175}&\frac{92696}{467775}&-\frac{61623938}{70945875}\\
&&-\frac{524}{2835}&-\frac{1618}{14175}&\frac{612536}{467775}&\frac{427003576}{1915538625}\\
&&&-\frac{5933}{14175}&-\frac{8324}{66825}&\frac{427770788}{212837625}\\
&&&&-\frac{320044}{467775}&-\frac{9153184}{70945875}\\
&&&&&-\frac{1978771378}{1915538625}
\end{bmatrix}
;\end{equation}
\begin{equation}\label{C-theta-xi}
\mathsf C_{\theta\xi}=
\begin{bmatrix}
-\frac{2}{3}&\frac{4}{45}&-\frac{158}{315}&-\frac{2102}{14175}&\frac{109042}{467775}&\frac{216932}{2627625}\\
&\frac{16}{45}&-\frac{16}{945}&\frac{934}{14175}&-\frac{7256}{155925}&\frac{117952358}{638512875}\\
&&-\frac{232}{2835}&\frac{922}{14175}&-\frac{25286}{66825}&-\frac{7391576}{54729675}\\
&&&\frac{719}{4725}&\frac{268}{18711}&-\frac{67048172}{638512875}\\
&&&&\frac{14354}{467775}&\frac{46774256}{638512875}\\
&&&&&\frac{253129538}{1915538625}
\end{bmatrix}
;\end{equation}
\begin{equation}\label{C-xi-mu}
\mathsf C_{\xi\mu}=\ifsqueeze\arraycolsep=3pt\fi
\begin{bmatrix}
\frac{1}{6}&-\frac{4}{45}&-\frac{817}{10080}&\frac{1297}{18900}&\frac{7764059}{239500800}&-\frac{9292991}{302702400}\\
&\frac{49}{720}&-\frac{2}{35}&-\frac{29609}{453600}&\frac{35474}{467775}&\frac{36019108271}{871782912000}\\
&&\frac{4463}{90720}&-\frac{2917}{56700}&-\frac{4306823}{59875200}&\frac{3026004511}{30648618000}\\
&&&\frac{331799}{7257600}&-\frac{102293}{1871100}&-\frac{368661577}{4036032000}\\
&&&&\frac{11744233}{239500800}&-\frac{875457073}{13621608000}\\
&&&&&\frac{453002260127}{7846046208000}
\end{bmatrix}
;\end{equation}
\begin{equation}\label{C-mu-xi}
\mathsf C_{\mu\xi}=\ifsqueeze\arraycolsep=0.5pt\fi
\begin{bmatrix}
-\frac{1}{6}&\frac{4}{45}&\frac{121}{1680}&-\frac{1609}{28350}&-\frac{384229}{14968800}&\frac{12674323}{851350500}\\
&-\frac{29}{720}&\frac{26}{945}&\frac{16463}{453600}&-\frac{431}{17325}&-\frac{31621753811}{1307674368000}\\
&&-\frac{1003}{45360}&\frac{449}{28350}&\frac{3746047}{119750400}&-\frac{32844781}{1751349600}\\
&&&-\frac{40457}{2419200}&\frac{629}{53460}&\frac{10650637121}{326918592000}\\
&&&&-\frac{1800439}{119750400}&\frac{205072597}{20432412000}\\
&&&&&-\frac{59109051671}{3923023104000}
\end{bmatrix}
;\end{equation}
\begin{equation}\label{C-xi-chi}
\mathsf C_{\xi\chi}=
\begin{bmatrix}
\frac{2}{3}&-\frac{34}{45}&\frac{46}{315}&\frac{2458}{4725}&-\frac{55222}{93555}&\frac{2706758}{42567525}\\
&\frac{19}{45}&-\frac{256}{315}&\frac{3413}{14175}&\frac{516944}{467775}&-\frac{340492279}{212837625}\\
&&\frac{248}{567}&-\frac{15958}{14175}&\frac{206834}{467775}&\frac{4430783356}{1915538625}\\
&&&\frac{16049}{28350}&-\frac{832976}{467775}&\frac{62016436}{70945875}\\
&&&&\frac{15602}{18711}&-\frac{651151712}{212837625}\\
&&&&&\frac{2561772812}{1915538625}
\end{bmatrix}
;\end{equation}
\begin{equation}\label{C-chi-xi}
\mathsf C_{\chi\xi}=
\begin{bmatrix}
-\frac{2}{3}&\frac{34}{45}&-\frac{88}{315}&-\frac{2312}{14175}&\frac{27128}{93555}&-\frac{55271278}{212837625}\\
&\frac{1}{45}&-\frac{184}{945}&\frac{6079}{14175}&-\frac{65864}{155925}&\frac{106691108}{638512875}\\
&&-\frac{106}{2835}&\frac{772}{14175}&-\frac{14246}{467775}&\frac{5921152}{54729675}\\
&&&-\frac{167}{9450}&-\frac{5312}{467775}&\frac{75594328}{638512875}\\
&&&&-\frac{248}{13365}&\frac{2837636}{638512875}\\
&&&&&-\frac{34761247}{1915538625}
\end{bmatrix}
.\end{equation}}

The transformation matrix to convert the coefficients for the expansions
in $n$ into expansions in $e^2$ is
{\ifsqueeze\small\fi
\begin{equation}\label{T-e2}
\mathsf T(n,e^2)=
\begin{bmatrix}
\frac{1}{4}&\frac{1}{8}&\frac{5}{64}&\frac{7}{128}&\frac{21}{512}&\frac{33}{1024}\\
&\frac{1}{16}&\frac{1}{16}&\frac{7}{128}&\frac{3}{64}&\frac{165}{4096}\\
&&\frac{1}{64}&\frac{3}{128}&\frac{27}{1024}&\frac{55}{2048}\\
&&&\frac{1}{256}&\frac{1}{128}&\frac{11}{1024}\\
&&&&\frac{1}{1024}&\frac{5}{2048}\\
&&&&&\frac{1}{4096}
\end{bmatrix}
.
\end{equation}}%
For example, the coefficients for converting $\phi$ to $\mu$ as an
expansion in $e^2$ are
{\ifsqueeze\small\fi
\begin{equation}\label{C-mu-phi-e2}
\mathsf C_{\mu\phi}\cdot\mathsf T(n,e^2)=\ifsqueeze\arraycolsep=1pt\fi
\begin{bmatrix}
-\frac{3}{8}&-\frac{3}{16}&-\frac{111}{1024}&-\frac{141}{2048}&-\frac{1533}{32768}&-\frac{2193}{65536}\\
&\frac{15}{256}&\frac{15}{256}&\frac{405}{8192}&\frac{165}{4096}&\frac{274695}{8388608}\\
&&-\frac{35}{3072}&-\frac{35}{2048}&-\frac{4935}{262144}&-\frac{29225}{1572864}\\
&&&\frac{315}{131072}&\frac{315}{65536}&\frac{13671}{2097152}\\
&&&&-\frac{693}{1310720}&-\frac{693}{524288}\\
&&&&&\frac{1001}{8388608}
\end{bmatrix}
.
\end{equation}}%
The inverse of $\mathsf T(n,e^2)$ is
{\ifsqueeze\small\fi
\begin{equation}\label{T-e2-inv}
\mathsf T(e^2,n)=
\begin{bmatrix}
4&-8&12&-16&20&-24\\
&16&-64&160&-320&560\\
&&64&-384&1344&-3584\\
&&&256&-2048&9216\\
&&&&1024&-10240\\
&&&&&4096
\end{bmatrix}
;
\end{equation}}%
this can be used to transform an expansion in $e^2$ into an expansion in $n$.
\null\vfill
\end{document}